\documentclass[aps,preprint,eqsecnum,showpacs,nofootinbib]{revtex4}
\usepackage{epsfig}
\begin{document}

\title{Next-to-leading order QCD predictions for
associated production of top squarks and charginos at the CERN
LHC}

\author{Li Gang Jin, Chong Sheng Li\footnote {E-mail: csli@pku.edu.cn}, and Jian Jun Liu}
\affiliation{ Department of Physics, Peking University, Beijing
100871, P.R. China}
%
%\date{\today}

\begin{abstract}
We present the calculations of the complete next-to-leading order
(NLO) inclusive total cross sections for the associated production
processes $pp\rightarrow \tilde{t}_i\tilde{\chi}_k^-+X$ in the
Minimal Supersymmetric Standard Model at the CERN Large Hadron
Collider. Our calculations show that the total cross sections for
the $\tilde{t}_1\tilde{\chi}_1^-$ production for the lighter top
squark masses in the region 100 GeV $< m_{\tilde{t}_1}<$ 160 GeV
can reach 1 pb in the favorable parameter space allowed by the
current precise experiments, and in other cases the total cross
sections generally vary from $10$ fb to several hundred fb except
both $m_{\tilde{t}_1}>$ 500 GeV and the
$\tilde{t}_2\tilde{\chi}_2^-$ production channel. Moreover, we
find that the NLO QCD corrections in general enhance the leading
order total cross sections significantly, and vastly reduce the
dependence of the total cross sections on the
renormalization/factorization scale, which leads to increased
confidence in predictions based on these results.
\end{abstract}

\maketitle

\section{Introduction}

The CERN Large Hadron Collider (LHC), with $\sqrt{S}=14$ TeV and a
luminosity of 100 ${\rm fb^{-1}}$ per year \cite{lhc}, will be a
wonderful machine for discovering new physics. In so many new
physical models, the Minimal Supersymmetric Standard Model (MSSM)
\cite{nilles} is one of the most attractive models for the
theorists and the high energy experimenters, and searching for
supersymmetric (SUSY) particles, as a direct experimental
evidence, is one of the prime objectives at the LHC. Therefore, a
good understanding of theoretical predictions of the cross
sections for the production of the SUSY particles is important.

The cross sections for the production of squarks and gluinos were
calculated at the Born level already many years ago \cite{squark}.
To date, the productions of gluinos and squarks
\cite{beenakker2,beenakker6}, top squarks \cite{beenakker3},
sleptons \cite{beenakker4,baer} and gauginos \cite{beenakker4} at
the hadron colliders have been studied. Besides the leading order
(LO) results, these calculations also have included the
next-to-leading order (NLO) corrections to improve the accuracy of
the theoretical predictions and reduce the uncertainty from the
dependence of results on renormalization/factorization scale. It
was recently pointed out in Ref.~\cite{berger} that the associated
production of a gaugino ($\tilde{\chi}$) with a gluino
($\tilde{g}$) or with a squark ($\tilde{q}$) is potentially a very
important production mechanism, since the mass spectrum favors
much lighter masses for the color-neutral charginos and
neutralinos than for the colored squarks and gluinos in popular
models of SUSY breaking. And a detailed NLO QCD calculation of the
associated production of a gluino and a gaugino at the Tevatron
and the LHC has been given in Ref.~\cite{berger}. However, the
processes of the associated gaugino and squark productions have
not been calculated at LO and NLO yet.

In this paper, we present the complete NLO QCD (including SUSY
QCD) calculation for the cross sections of the associated
production of top squarks (stops) and charginos at the LHC.
Similar to $pp \rightarrow gb \rightarrow t H^-$ \cite{charged},
which is expected to be a dominate process for the charged Higgs
boson production at the LHC, the associated production $pp
\rightarrow gb \rightarrow \tilde{t}_i\tilde{\chi}_k^-$ may be
also the dominate process for single top squark or chargino
production at the LHC. This is due to the following reasons:
first, the large top quark mass in stop mass matrix can lead to
strong mixing, and induce large mass difference between the
lighter mass eigenstate and the heavier one, which means that the
phase space for the lighter stop will be great and benefit its
production; second, besides containing a strong QCD coupling
between the incoming partons, this process also includes an
enhanced effects from the Yukawa coupling in the vertex
$b-\tilde{t}_i-\tilde{\chi}_k^-$ of the final states. For
simplicity, in this paper we neglect the bottom quark mass except
in the Yukawa coupling. Such approximations are valid in all
diagrams, in which the bottom quark appears as an initial state
parton, according to the simplified Aivazis-Collins-Olness-Tung
(ACOT) scheme \cite{acot}. However, it was pointed out in
Ref.~\cite{nbpdf} that the approximations of the hard process
kinematics and the introduction of conventional bottom quark
densities will give rise to sizable bottom quark mass and
kinematical phase space effects, and may overestimate the
inclusive cross section, for example, in the processes of
$b\bar{b}\rightarrow H$ \cite{bbbH} and $bg\rightarrow bH$
\cite{gbbH}. Very recently, it is shown in Ref.~\cite{bpdf1} that
the bottom parton approach is still valid if we choose the
factorization scale below the average final state mass: $\mu_f\sim
C m_{\rm av}\equiv C (m_{\tilde{t}_i}+m_{\tilde{\chi}_k^-})/2$
with $C\sim (1/4,...,1/3)$. Thus, in this paper, we choose
$\mu_f=m_{\rm av}/3$ when we use the bottom parton approximation.

The arrangement of this paper is as follows. In Sec. II we show
the LO results and define the notations. In Sec. III we present
the details of the calculations of both the virtual and real parts
of the NLO QCD corrections. In Sec. IV by a detailed numerical
analysis we present the predictions for inclusive and differential
cross sections at the LHC. The lengthy analytic expressions are
summarized in Appendixes A and B.

\section{Leading Order Production of Stops and Charginos}

\begin{figure}[h]
\centerline{\epsfig{file=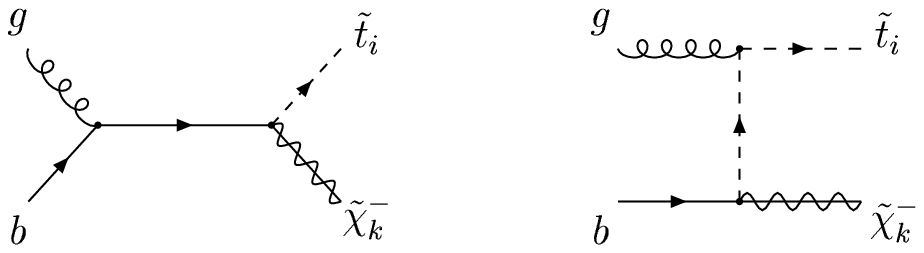, width=250pt}}
\caption[]{Leading order feynman diagrams for $gb\rightarrow
\tilde{t}_i \tilde{\chi}_k^-$. \label{fyntree}}
\end{figure}

The partonic process of the LO associated production of stops and
charginos is $g(p_a) b(p_b) \rightarrow \tilde{t}_i(p_1)
\tilde{\chi}_k^-(p_2)$, and the related Feynman diagrams are shown
in Fig.~\ref{fyntree}. In order to simplify our expressions, we
introduce the following Mandelstam variables:
\begin{eqnarray}
&& s=(p_a+p_b)^2, \ \ s_\Delta=s-m_{\tilde{t}_i}^2
-m_{\tilde{\chi}_k^-}^2, \nonumber \\
&& t=(p_a-p_1)^2, \ \ \ t_1=t-m_{\tilde{t}_i}^2, \ \ \
t_2=t-m_{\tilde{\chi}_k^-}^2, \nonumber \\
&& u=(p_a-p_2)^2, \ \  u_1=u- m_{\tilde{t}_i}^2, \ \ u_2=u-
m_{\tilde{\chi}_k^-}^2.
\end{eqnarray}
After the $n$-dimensional phase space integration, the LO partonic
differential cross sections are given by ($n=4 -2 \epsilon$)
\begin{eqnarray}
&& \frac{d^2 \hat{\sigma}_{ik}^B}{dt_2 du_2} = \frac{\pi
S_{\epsilon}}{s^2\Gamma(1-\epsilon)} (\frac{t_2u_2 - s
m_{\tilde{\chi}_k^-}^2}{\mu_r^2s})^{-\epsilon} \Theta(t_2u_2 -s
m_{\tilde{\chi}_k^-}^2) \Theta[s- (m_{\tilde{t}_i}
+m_{\tilde{\chi}_k^-})^2] \nonumber \\ && \hspace{1.8cm} \times
\delta (s+ t +u - m_{\tilde{t}_i}^2 -m_{\tilde{\chi}_k^-}^2)
\overline{|M_{ik}^B|}^2
\end{eqnarray}
with
\begin{eqnarray}
\overline{|M^B_{ik}|}^2 =\frac{g^2g_s^2}{12(1 -\epsilon)}
(|l^{\tilde{t}}_{ik}|^2+|k^{\tilde{t}}_{ik}|^2)
(-\frac{t_2+(1-\epsilon)u_2}{s}
+\frac{ss_{\Delta}-u_1(u_2+2t_2)}{st_1}
+\frac{2t_2m_{\tilde{t}_i}^2}{t_1^2}),
\end{eqnarray}
where $S_{\epsilon} = (4\pi)^{-2 +\epsilon}$, and the indices
$(i,k)$ label the outgoing particles ($\tilde{t}_i,
\tilde{\chi}_k^-$). The scale parameter $\mu_r$ is introduced to
provide the correct mass dimension for the coupling constant in
$n$-dimensions. $\overline{|M_{ik}^{\rm B}|}^2$ is the LO squared
matrix element, which has been summed the colors and spins of the
outgoing particles, and averaged over the colors and spins of the
incoming ones. $l^{\tilde{t}}_{ik}$ and $k^{\tilde{t}}_{ik}$ are
the left- and right- handed coupling constants of the vertex
$b-\tilde{t}_i-\tilde{\chi}_k^-$, respectively, and are defined
together with $l^{\tilde{b}}_{ik}$ and $k^{\tilde{b}}_{ik}$ in the
vertex $t-\tilde{b}_i-\tilde{\chi}_k^+$, which is involved in the
virtual corrections, as follows:
\begin{eqnarray}
\left(\begin{array}{c} l^{\tilde{t}}_{ik} \\ l^{\tilde{b}}_{ik}
\end{array} \right)=-\left (\begin{array}{c}
R^{\tilde{t}}_{i1}V_{k1}^{\ast} \\ R^{\tilde{b}}_{i1}U_{k1}^{\ast}
\end{array} \right) +\left (\begin{array}{c} Y_t
R_{i2}^{\tilde{t}}V_{k2}^{\ast} \\
Y_bR_{i2}^{\tilde{b}}U_{k2}^{\ast} \end{array} \right), \ \
\ \ \left(\begin{array}{c} k^{\tilde{t}}_{ik} \\
k^{\tilde{b}}_{ik}
\end{array} \right)=\left (\begin{array}{c} Y_b
R^{\tilde{t}}_{i1}U_{k2} \\ Y_t R^{\tilde{b}}_{i1}V_{k2}
\end{array} \right)
\end{eqnarray}
with
\begin{eqnarray}
Y_t \equiv \frac{m_t}{\sqrt{2}m_W\sin\beta} \equiv \frac{h_t}{g} ,
\ \ \ Y_b\equiv \frac{m_b}{\sqrt{2}m_W\cos\beta}  \equiv
\frac{h_b}{g}.
\end{eqnarray}
Here the angle $\beta$ is defined by $\tan \beta\equiv v_2/v_1$,
the ratio of the vacuum expectation values of the two Higgs
doublets. Matrices $U$ and $V$ are the chargino transformation
matrices from interaction to mass eigenstates defined in
Ref.~\cite{gunion}. Matrix $R^{\tilde{q}}$ $(q=t,b)$ is defined to
transform the current eigenstates $\tilde{q}_L$ and $\tilde{q}_R$
to the mass eigenstates $\tilde{q}_1$ and $\tilde{q}_2$:
\begin{equation}
\left(\begin{array}{c} \tilde{q}_1 \\ \tilde{q}_2 \end{array}
\right)= R^{\tilde{q}}\left(\begin{array}{c} \tilde{q}_L \\
\tilde{q}_R \end{array} \right), \ \ \ \ \
R^{\tilde{q}}=\left(\begin{array}{cc} \cos\theta_{\tilde{q}} &
\sin\theta_{\tilde{q}} \\ -\sin\theta_{\tilde{q}} &
\cos\theta_{\tilde{q}}
\end{array} \right)
\end{equation}
with $0 \leq \theta_{\tilde{q}} < \pi$ by convention.
Correspondingly, the mass eigenvalues $m_{\tilde{q}_1}$ and
$m_{\tilde{q}_2}$ (with $m_{\tilde{q}_1}\leq m_{\tilde{q}_2}$) are
given by
\begin{eqnarray}
\left(\begin{array}{cc} m_{\tilde{q}_1}^2 & 0 \\ 0 &
m_{\tilde{q}_2}^2 \end{array} \right)=R^{\tilde{q}}
M_{\tilde{q}}^2 (R^{\tilde{q}})^\dag, \ \ \ \ \
M_{\tilde{q}}^2=\left(\begin{array}{cc} m_{\tilde{q}_L}^2 & a_qm_q
\\ a_qm_q & m_{\tilde{q}_R}^2 \end{array} \right)
\end{eqnarray}
with
\begin{eqnarray}
m^2_{\tilde{q}_L} &=& M^2_{\tilde{Q}} +m_q^2
+m_Z^2\cos2\beta(I_{3L}^q -e_q\sin^2\theta_W), \\
m^2_{\tilde{q}_R} &=& M^2_{\{\tilde{U},\tilde{D}\}} +m_q^2
+m_Z^2\cos2\beta e_q\sin^2\theta_W, \\
a_q &=& A_q -\mu\{\cot\beta, \tan\beta\}
\end{eqnarray}
for \{up, down\} type squarks. Here $M_{\tilde{q}}^2$ is the
squark mass matrix. $M_{\tilde{Q},\tilde{U},\tilde{D}}$ and
$A_{t,b}$ are soft SUSY-breaking parameters and $\mu$ is the Higgs
mixing parameter in the superpotential. $I_{3L}^q$ and $e_q$ are
the third component of the weak isospin and the electric charge of
the quark $q$, respectively.

The LO total cross section at the LHC is obtained by convoluting
the partonic cross section with the parton distribution functions
(PDFs) $G_{b,g/p}$ in the proton:
\begin{equation}
\sigma^B_{ik}=\int dx_1dx_2 [G_{b/p}(x_1,\mu_f)G_{g/p}(x_2,\mu_f)+
(x_1\leftrightarrow x_2)]\hat{\sigma}^{B}_{ik},
\end{equation}
where $\mu_f$ is the factorization scale.

\section{Next-to-Leading Order Calculations}
The NLO contributions to the associated production of stops and
charginos can be separated into the virtual corrections arising
from loop diagrams of colored particles and the real corrections
arising from the radiations of a real gluon or a massless
(anti)quark.

\subsection{Virtual Corrections}
\begin{figure}
\centerline{\epsfig{file=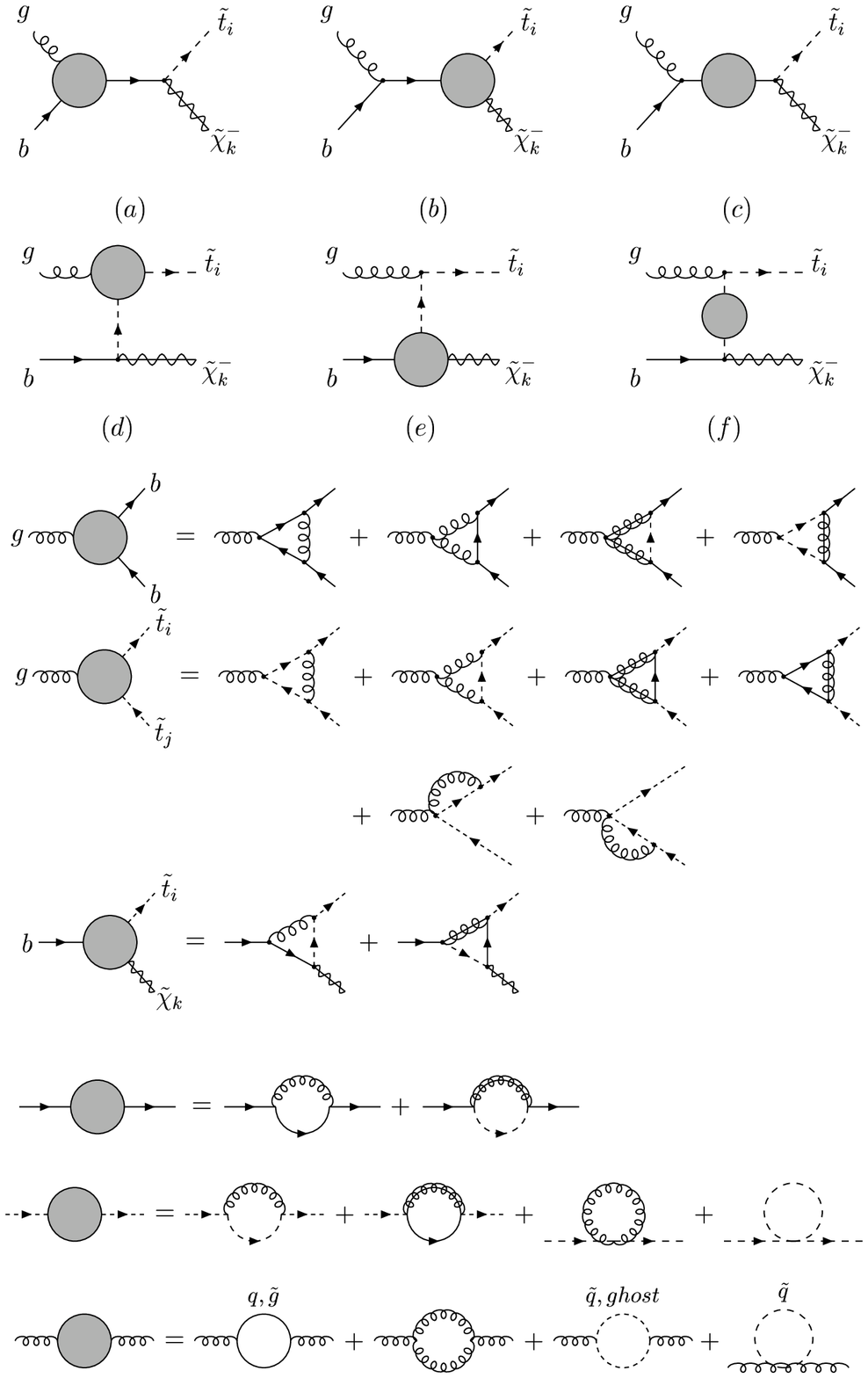, width=350pt}} \vspace{0.8cm}
\caption[]{Virtual one-loop Feynman diagrams including self-energy
and vertex corrections for $gb\rightarrow \tilde{t}_i
\tilde{\chi}_k^-$.\label{fynvertex}}
\end{figure}

\begin{figure}
\centerline{\epsfig{file=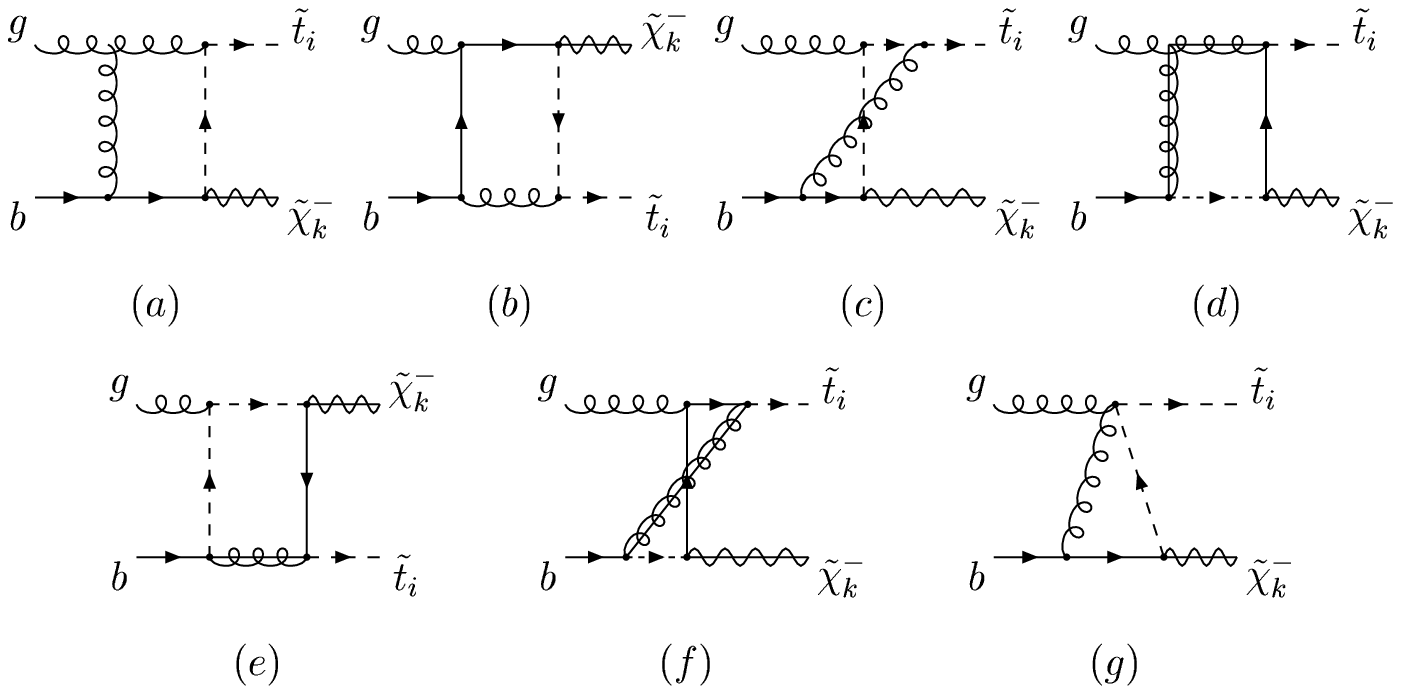, width=350pt}} \caption[]{Box
Feynman diagrams for $gb\rightarrow \tilde{t}_i \tilde{\chi}_k^-$.
\label{fynbox}}
\end{figure}

The virtual corrections to $gb\rightarrow
\tilde{t}_i\tilde{\chi}_k^-$ arise from the Feynman diagrams shown
in Fig.~\ref{fynvertex} and \ref{fynbox}, which consist of
self-energy, vertex and box diagrams. We carried out the
calculation in t'Hooft-Feynman gauge and used the dimensional
regularization in $n=4 -2\epsilon$ dimensions to regularize the
ultraviolet (UV), soft infrared and collinear divergences in the
virtual loop corrections. However, this method violates the
supersymmetry. In order to restore the supersymmetry the simplest
procedure is through finite shifts in the
quark-squark-gluino(chargino) couplings \cite{beenakker5}:
\begin{eqnarray}
&& \hat{g}_s = g_s [1 +\frac{\alpha_s}{8\pi}(\frac{4}{3}N - C_F)],
\nonumber \\
&& \hat{e}=e(1- \frac{\alpha_s}{8\pi} C_F), \nonumber \\
&& \hat{h}_{t,b}=h_{t,b}( 1- \frac{3\alpha_s}{8\pi}C_F)
\end{eqnarray}
with $N=3$ and $C_F=4/3$. Since we have not such a coupling as
$\tilde{g}-q-\tilde{q}$ at the tree-level, the first shift is not
used in our calculations, and we just consider the latter two
shifts. As for the Dirac matrix $\gamma_5$, we deal with it using
the ``naive" scheme, in which the $\gamma_5$-matrix anticommutes
with the other $\gamma_{\mu}$-matrices. This is a legitimate
procedure at the one-loop level for anomaly-free theories
\cite{gamma5}.

The virtual corrections to the differential cross section can be
expressed as
\begin{eqnarray}
&& \frac{d^2 \hat{\sigma}_{ik}^V}{dt_2 du_2} = \frac{\pi
S_{\epsilon}}{s^2\Gamma(1-\epsilon)} (\frac{t_2u_2 - s
m_{\tilde{\chi}_k^-}^2}{\mu_r^2s})^{-\epsilon} \Theta(t_2u_2 -s
m_{\tilde{\chi}_k^-}^2) \Theta[s- (m_{\tilde{t}_i}
+m_{\tilde{\chi}_k^-})^2] \nonumber \\ && \hspace{1.8cm} \times
\delta (s+ t +u - m_{\tilde{t}_i}^2 -m_{\tilde{\chi}_k^-}^2) \ 2 \
{\rm Re} (\overline{M^V_{ik}M_{ik}^{B\ast}}),
\end{eqnarray}
where $M^V_{ik}$ is the renormalized amplitude, which can be
separated into two parts:
\begin{equation}
M^{V}_{ik}=M^{unren}_{ik}+M^{con}_{ik}.
\end{equation}
Here $M^{unren}_{ik}$ contains the self-energy, vertex and box
corrections, and $M^{con}_{ik}$ is the corresponding counterterm.
Moreover, $M^{unren}_{ik}$ can be written as
\begin{equation}
M_{unren}=\sum_{\alpha=a}^f M_{ik}^\alpha + \sum_{\beta=a}^g
M_{ik}^{{\rm box} (\beta)},
\end{equation}
where $\alpha$ and $\beta$ denote the corresponding diagram
indexes in Fig.~\ref{fynvertex} and Fig.~\ref{fynbox},
respectively. We express each term further as
\begin{eqnarray}
&& M_{ik}^\alpha=C_s \sum_{l=1}^{12} f_l^{\alpha} M_l,
\label{fvertex}
\\
&& M_{ik}^{box (\beta)}=C_s \sum_{l=1}^{12} f_l^{box (\beta)} M_l,
\label{fbox}
\end{eqnarray}
where $C_s = i gg_s^3/(16\pi^2)$, $f_l^\alpha$ and
$f_l^{box(\beta)}$ are the form factors, which are given
explicitly in Appendix A, and $M_l$ are the standard matrix
elements defined as
\begin{eqnarray}
&& M_{1,2}=v^T(p_2)C^{-1}T^a \not{\varepsilon}(p_a)P_{L,R}\
u(p_b), \nonumber
\\ && M_{3,4}=v^T(p_2)C^{-1}T^a \not{p}_a
\not{\varepsilon}(p_a)P_{L,R}\ u(p_b), \nonumber
\\ && M_{5,6}=v^T(p_2)C^{-1}T^a P_{L,R}\ u(p_b)p_b \cdot \varepsilon(p_a), \nonumber
\\ && M_{7,8}=v^T(p_2)C^{-1}T^a P_{L,R}\ u(p_b)p_2 \cdot \varepsilon(p_a), \nonumber
\\ && M_{9,10}=v^T(p_2)C^{-1}T^a \not{p}_aP_{L,R}\ u(p_b)p_b \cdot
\varepsilon(p_a), \nonumber
\\ && M_{11,12}=v^T(p_2)C^{-1}T^a \not{p}_aP_{L,R}\ u(p_b)p_2 \cdot
\varepsilon(p_a).
\end{eqnarray}
Here $C$ is the charge conjugation operator, and $T^a$ is the
$SU(3)$ color matrix.

$M^{con}_{ik}$ is separated into $M^{con(s)}_{ik}$ and
$M^{con(t)}_{ik}$, i.e. the counterterms for $s$ and $t$ channels,
respectively, which are given by ($j\neq i$)
\begin{eqnarray*}
&& M^{con}_{ik}= M_{ik}^{con(s)} + M_{ik}^{con(t)},
\\
&& M_{ik}^{con(s)} = -\frac{igg_s}{s} \{ [l_{ik}^{\tilde{t}}
\Delta +l_{jk}^{\tilde{t}} \delta Z_{ji}^{\tilde{t}} +\delta
l_{ik}^{\tilde{t}}](M_3 +2M_5) +[k_{ik}^{\tilde{t}} \Delta
+k_{jk}^{\tilde{t}} \delta Z_{ji}^{\tilde{t}} +\delta
k_{ik}^{\tilde{t}}](M_4 +2M_6) \},
\\
&& M_{ik}^{con(t)} = -i2gg_s \{[\frac{l_{ik}^{\tilde{t}}} {t_1}
(\Delta +\frac{\delta m_{\tilde{t}_i}^2 }{t_1}) +\frac{\delta
l_{ik}^{\tilde{t}}}{t_1} + \frac{l_{jk}^{\tilde{t}}\delta
Z_{ji}^{\tilde{t}} }{t-m_{\tilde{t}_j}^2}] (M_5 - M_7)
+[\frac{k_{ik}^{\tilde{t}}} {t_1} (\Delta +\frac{\delta
m_{\tilde{t}_i}^2 }{t_1})
\\ && \hspace{1.8cm}
+\frac{\delta k_{ik}^{\tilde{t}}}{t_1} +
\frac{k_{jk}^{\tilde{t}} \delta
Z_{ji}^{\tilde{t}}}{t-m_{\tilde{t}_j}^2} ] (M_6 - M_8) \},
\end{eqnarray*}
with
\begin{eqnarray*}
&& \Delta=\frac{\delta g_s}{g_s} + \frac{1}{2} \delta Z_g +
\frac{1}{2} \delta Z_b + \frac{1}{2} \delta Z_{ii}^{\tilde{t}},
\\
&& \delta l_{ik}^{\tilde{t}} = R_{i2}^{\tilde{t}} V_{k1}^{\ast}
\delta\theta^{\tilde{t}} + Y_t V_{k2}^{\ast} (R_{i2}^{\tilde{t}}
\frac{\delta m_t }{m_t} + R_{i1}^{\tilde{t}}
\delta\theta^{\tilde{t}}),
\\
&& \delta k_{ik}^{\tilde{t}} = Y_b U_{k2}(R_{i1}^{\tilde{t}}
\frac{\delta m_b }{m_b} - R_{i2}^{\tilde{t}}
\delta\theta^{\tilde{t}}).
\end{eqnarray*}
where $\delta m^2_{\tilde{t}_i}$, $\delta m_q \ (q=t,b)$, $\delta
Z_b$, $\delta Z_g$, $\delta Z_{ii}^{\tilde{t}}$, $\delta
Z_{ij}^{\tilde{t}}$ and $\delta \theta_{\tilde{t}}$ are the
renormalization constants, which are fixed by the on-shell
renormalization scheme \cite{onmass}, and can be written as
(assuming $m_{\tilde{b}_1}=m_{\tilde{b}_2}$)
\begin{eqnarray*}
&& \delta m_{\tilde{t}_i}^2= -\frac{\alpha_s}{4\pi} C_F \{4
m_{\tilde{t}_i}^2 B_0(m_{\tilde{t}_i}^2,m_{\tilde{t}_i}^2,0)
-A_0(m_{\tilde{t}_i}^2) +4A_0(m_t^2) +4[m_{\tilde{t}_i}^2B_1
+m_{\tilde{g}}^2B_0
\\ && \hspace{1.0cm}
-2m_tm_{\tilde{g}} R_{i1}^{\tilde{t}}R_{i2}^{\tilde{t}}
B_0](m_{\tilde{t}_i}^2,m_{\tilde{g}}^2,m_t^2)\},
\\
&& \frac{\delta m_q}{m_q} =  -\frac{\alpha_s}{4\pi} C_F \{(4B_0
+2B_1)(m_q^2,m_q^2,0) -1 + \sum_{i=1}^2 [B_1
-\frac{m_{\tilde{g}}}{m_q} \sin2\theta_{\tilde{q}} (-1)^i
B_0](m_q^2,m_{\tilde{g}}^2,m_{\tilde{q}_i}^2)\},
\\
&& \delta Z_b= \frac{\alpha_s}{2\pi}C_F
B_1(0,m_{\tilde{g}}^2,m_{\tilde{b}_1}^2),
\\
&& \delta Z_g= -\frac{\alpha_s}{4\pi} [\frac{2}{3}(B_0
+2m_t^2B_0')(0,m_t^2,m_t^2) +2(B_0 +2m_{\tilde{g}}^2B_0'
)(0,m_{\tilde{g}}^2,m_{\tilde{g}}^2)
\\ && \hspace{1.0cm}
+\frac{2}{3}\sum_{q,i}(\frac{1}{4}B_0 -m_{\tilde{q}_i}^2B_0'
)(0,m_{\tilde{q}_i}^2,m_{\tilde{q}_i}^2) -\frac{7}{9}],
\\
&& \delta Z_{ii}^{\tilde{t}} =\frac{\alpha_s}{2\pi} C_F \{(B_0 + 2
m_{\tilde{t}_i}^2 B_0')(m_{\tilde{t}_i}^2,m_{\tilde{t}_i}^2,0)
-[B_0 +(m_{\tilde{t}_i}^2 -m_{\tilde{g}}^2 -m_t^2) B_0'
\\ && \hspace{1.0cm}
+4 m_tm_{\tilde{g}} R_{i1}^{\tilde{t}} R_{i2}^{\tilde{t}}
B_0'](m_{\tilde{t}_i}^2,m_{\tilde{g}}^2 ,m_t^2) \},
\\
&& \delta Z_{ij}^{\tilde{t}} = \frac{{\rm Re}
\Sigma_{ij}^{\tilde{t}}(m_{\tilde{t}_j}^2)}{m_{\tilde{t}_i}^2
-m_{\tilde{t}_j}^2},
\\
&& \delta\theta_{\tilde{t}} =\frac{{\rm
Re}[\Sigma_{12}^{\tilde{t}}(m_{\tilde{t}_1}^2)
+\Sigma_{12}^{\tilde{t}}(m_{\tilde{t}_2}^2)]}{2(m_{\tilde{t}_1}^2
-m_{\tilde{t}_2}^2)},
\end{eqnarray*}
where $B_0'=\partial B_0/\partial p^2$, $A_0$ and $B_i$ are the
Passarino-Veltman one-point and two-point integrals, respectively,
which are defined similar to Ref.~\cite{denner} except that we
take internal masses squared as arguments. Here the counterterm
$\delta\theta_{\tilde{t}}$ for the stop mixing angle is fixed as
Ref.~\cite{theta}, and
\begin{eqnarray*}
&& \Sigma_{ij}^{\tilde{t}} =\frac{\alpha_s}{4\pi} \frac{2}{3}
\{\sin4\theta_{\tilde{t}} [A_0(m_{\tilde{t}_2}^2)
-A_0(m_{\tilde{t}_1}^2)] +8m_tm_{\tilde{g}}\cos2\theta_{\tilde{t}}
B_0(m_{\tilde{t}_j}^2,m_{\tilde{g}}^2,m_t^2)\}.
\end{eqnarray*}
In $\delta Z_{ii}^{\tilde{t}}$,
$B_0'(m_{\tilde{t}_i}^2,m_{\tilde{t}_i}^2,0)$ contains infrared
divergences, and can be expressed as
\begin{eqnarray*}
B_0'(m_{\tilde{t}_i}^2,m_{\tilde{t}_i}^2,0)=
-\frac{1}{2m_{\tilde{t}_i}^2}(\frac{4\pi\mu_r^2}
{m_{\tilde{t}_i}^2})^{\epsilon} \Gamma(1+\epsilon)
(\frac{1}{\epsilon} +2).
\end{eqnarray*}
Moreover, the QCD coupling constant $g_s$ is renormalized in the
modified minimal subtraction ($\overline{MS}$) scheme except that
the divergences associated with the top quark and colored SUSY
particle loops are subtracted at zero momentum \cite{subtract}.
Assuming that the scalar partners of the $n_f=5$ light quark
flavors have a common mass $m_{\tilde{q}}$, we obtain
\begin{eqnarray}
&& \frac{\delta g_s}{g_s}= -\frac{\alpha_s(\mu_r^2)}{4\pi}
[\frac{\beta_0}{2} \frac{1}{\bar{\epsilon}}
+\frac{N}{3}\ln\frac{m_{\tilde{g}}^2} {\mu_r^2}
+\frac{n_f}{6}\ln\frac{m_{\tilde{q}}^2}{\mu_r^2}
+\frac{1}{12}\ln\frac{m_{\tilde{t}_1}^2}{\mu_r^2}
+\frac{1}{12}\ln\frac{m_{\tilde{t}_2}^2}{\mu_r^2}
+\frac{1}{3}\ln\frac{m_t^2}{\mu_r^2}],
\end{eqnarray}
which agrees with one given in Ref.~\cite{beenakker2}. Here
$1/\bar{\epsilon}=1/\epsilon_{UV} -\gamma_E +\ln(4\pi)$, and
\begin{eqnarray}
&& \beta_0=(\frac{11}{3}N-\frac{2}{3}n_f) +[-\frac{2}{3}(N+1)
-\frac{1}{3}(n_f+1)]\equiv\beta^L_0 +\beta^H_0,
\end{eqnarray}
where $\beta^L_0$ includes the contributions from the gluon and
the quarks except the top quark, and $\beta^H_0$ contains the
contributions from the top quark and all colored SUSY particles.
The evolution of $g_s$ is determined by $\beta_0^L$. After
renormalization, $M^V_{ik}$ is UV-finite, but it still contains
the infrared (IR) divergences:
\begin{eqnarray}
M^V_{ik}|_{IR} =\frac{\alpha_s}{2\pi} \Gamma(1+\epsilon)
(\frac{4\pi\mu_r^2}{s})^\epsilon (\frac{A_2^V}{\epsilon^2}
+\frac{A_1^V}{\epsilon})M_{ik}^B,
\end{eqnarray}
where
\begin{eqnarray}
&& A_2^V=-\frac{13}{3}, \\
&& A_1^V=-\frac{43}{6} -\frac{4}{3}\ln\frac{s}{m_{\tilde{t}_i}^2}
+3\ln\frac{-t_1}{m_{\tilde{t}_i}^2} -\frac{1}{3}\ln\frac{-
u_1}{m_{\tilde{t}_i}^2}.
\end{eqnarray}
Here the infrared divergences include the soft infrared
divergences and the collinear infrared divergences. The soft
infrared divergences can be cancelled by adding the real
corrections, and the remaining collinear infrared divergences can
be absorbed into the redefinition of PDF \cite{altarelli}, which
will be discussed in the following subsections.

\subsection{Real Gluon Emission}
The Feynman diagrams of the real gluon emission process $g(p_a)b
(p_b)\rightarrow \tilde{t}_i (p_1) \tilde{\chi}^-_k (p_2)+g (p_3)$
are shown in Fig.4. Using the notations analogous to ones defined
in Ref.~\cite{beenakker1}, we express our results in terms of the
following invariants:
\begin{eqnarray}
&& s=(p_a +p_b)^2, \hspace{1.7cm} s_5=(p_1+p_2)^2, \hspace{1.5cm}
s_{5\Delta} =s_5 -\Delta_s, \nonumber
\\ &&
s_3=(p_2 +p_3)^2 -m_{\tilde{t}_i}^2, \ \ \ s_4=(p_1 +p_3)^2 -
m_{\tilde{t}_i}^2, \ \ \ s_{3\Delta}=s_3 -\Delta_t, \nonumber
\\ &&
t=(p_b -p_2)^2, \ \ \ t'=(p_b-p_3)^2, \ \ \ t_1=
t-m_{\tilde{t}_i}^2, \ \ \ t_2=t -m_{\tilde{\chi}_k^-}^2,
\nonumber
\\ &&
u=(p_a-p_2)^2, \ \ u'=(p_a -p_3)^2, \ \ u_1=u -m_{\tilde{t}_i}^2,
\ \ u_2=u -m_{\tilde{\chi}_k^-}^2, \nonumber
\\ &&
u_6=(p_b -p_1)^2 -m_{\tilde{t}_i}^2, \hspace{2.2cm} u_7=(p_a-
p_1)^2 -m_{\tilde{t}_i}^2,
\end{eqnarray}
where $\Delta_s=m_{\tilde{t}_i}^2 + m_{\tilde{\chi}_k^-}^2$ and
$\Delta_t=m_{\tilde{\chi}_k^-}^2 -m_{\tilde{t}_i}^2$.

The phase space integration for the real gluon emission will
produce infrared singularities, which can be either soft or
collinear and can be conveniently isolated by slicing the phase
space into different regions defined by suitable cutoffs. In this
paper, we use the two cutoff phase space slicing method
\cite{cutoff} to decompose the three-body phase space into three
regions.

First, the phase space can be separated into two regions by an
arbitrary small soft cutoff $\delta_s$, according to whether the
energy of the emitted gluon is soft, i.e. $E_3\leq
\delta_s\sqrt{s}/2$, or hard, i.e. $E_3> \delta_s\sqrt{s}/2$.
Correspondingly, the partonic real cross section can be written as
\begin{eqnarray}
\hat{\sigma}_{ik}^{R}= \hat{\sigma}_{ik}^{S}
+\hat{\sigma}_{ik}^{H},
\end{eqnarray}
where $\hat{\sigma}_{ik}^{S}$ and $\hat{\sigma}_{ik}^{H}$ are the
contributions from the soft and hard regions, respectively.
$\hat{\sigma}_{ik}^{S}$ contains all the soft infrared
divergences, which can be explicitly obtained after the
integration over the phase space of the emitted gluon. Second, to
isolate the remaining collinear divergences from
$\hat{\sigma}_{ik}^{H}$, we should introduce another arbitrary
small cutoff, called collinear cutoff $\delta_c$, to further split
the hard gluon phase space into two regions, according to whether
the Mandelstam variables satisfy the collinear condition $0< -t'\
(-u') < \delta_c s$ or not. Then we have
\begin{eqnarray}
\hat{\sigma}_{ik}^{H}= \hat{\sigma}_{ik}^{HC}+
\hat{\sigma}_{ik}^{\overline{HC}},
\end{eqnarray}
where the hard collinear part $\hat{\sigma}_{ik}^{HC}$ contains
the collinear divergences, which can be explicitly obtained after
the integration over the phase space of the emitted gluon. And the
hard non-collinear part $\hat{\sigma}_{ik}^{\overline{HC}}$ is
finite and can be numerically computed using standard Monte-Carlo
integration techniques \cite{monte}, and can be written as
\begin{eqnarray}
d\hat{\sigma}^{\overline{HC}}_{ik}=\frac{1}{2s}
\overline{|M_{ik}^{gb}|}^2 d\overline{\Gamma}_3. \label{nonHC}
\end{eqnarray}
Here $d\overline{\Gamma}_3$ is the hard non-collinear region of
the three-body phase space, and the explicit expressions of
$\overline{|M_{ik}^{gb}|}^2$ can be found in Appendix B. In order
to avoid introducing external ghost lines while summing over the
gluon helicities, we limit ourselves to the sum over the physical
polarizations of the gluons \cite{beenakker1}, i.e.
\begin{eqnarray}
P_i^{\mu\nu}=\sum_{T} \epsilon_T^\mu(k_i) \epsilon_T^\nu(k_i)=
-g^{\mu\nu} +\frac{n_i^\mu k_i^\nu +k_i^\mu n_i^\nu}{n_i\cdot k_i}
-\frac{n_i^2 k_i^\mu k_i^\nu}{(n_i\cdot k_i)^2},
\end{eqnarray}
where the index $i$ (=1,2) labels the two external gluons, and
$n_i\neq k_i$ is an arbitrary vector. This polarization sum obeys
the transversality relations
\begin{eqnarray}
k_{i\mu}P^{\mu\nu}=P^{\mu\nu} k_{i\nu}=
n_{i\mu}P^{\mu\nu}=P^{\mu\nu} n_{i\nu}=0.
\end{eqnarray}
In the next two subsections, we will discuss in detail the soft
and hard collinear gluon emission.

\subsubsection{Soft gluon emission}
In the limit that the energy of the emitted gluon becomes small,
i.e. $E_3\leq \delta_s\sqrt{s}/2$, the matrix element squared
$\overline{|M_{ik}^{R}|}^2$ can be simply factorized into the Born
matrix element squared times an eikonal factor $\Phi_{eik}$:
\begin{eqnarray}
\overline{|M_{ik}^{R}(gb\rightarrow \tilde{t}_i \tilde{\chi}^-_k
+g)|}^2 \stackrel{soft}{\rightarrow}
(4\pi\alpha_s\mu_r^{2\epsilon}) \overline{|M_{ik}^{B}|}^2
\Phi_{eik},
\end{eqnarray}
where the eikonal factor $\Phi_{eik}$ is given by
\begin{eqnarray}
\Phi_{eik}= N\frac{p_a\cdot p_1}{(p_a\cdot p_3)(p_1\cdot p_3)}
+N\frac{p_a\cdot p_b}{(p_b\cdot p_3)(p_a\cdot p_3)}
-C_F\frac{m_{\tilde{t}_i}^2}{(p_1\cdot p_3)^2}
-\frac{1}{N}\frac{p_b\cdot p_1}{(p_1\cdot p_3)(p_b\cdot p_3)}.
\end{eqnarray}
Moreover, the phase space in the soft limit can also be factorized
as
\begin{eqnarray}
d\Gamma_3(gb\rightarrow \tilde{t}_i \tilde{\chi}^-_k +g)
\stackrel{soft}{\rightarrow}  d\Gamma_2(gb\rightarrow \tilde{t}_i
\tilde{\chi}^-_k) dS,
\end{eqnarray}
where $dS$ is the integration over the phase space of the soft
gluon, which is given by \cite{cutoff}
\begin{eqnarray}
dS =\frac{1}{2(2\pi)^{3- 2\epsilon}} \int_0^{\delta_s \sqrt{s}/2}
dE_3 E_3^{1 -2\epsilon} d \Omega_{2-2 \epsilon}.
\end{eqnarray}
Then the parton level cross section in the soft region can be
expressed as
\begin{eqnarray}\label{soft}
&&\hat{\sigma}^S_{ik} =(4\pi\alpha_s\mu_r^{2\epsilon})\int
d\Gamma_2\overline{|M_{ik}^{B}|}^2 \int dS \Phi_{eik}.
\end{eqnarray}
Using the approach of Ref.~\cite{cutoff}, after the integration
over the soft gluon phase space, Eq.~(\ref{soft}) becomes
\begin{eqnarray}
&&\hat{\sigma}^S_{ik} =\hat{\sigma}^B [\frac{\alpha_s}{2\pi}
\frac{\Gamma(1-\epsilon)}{\Gamma(1-2\epsilon)}
(\frac{4\pi\mu_r^2}{s})^\epsilon] (\frac{A_2^s}{\epsilon^2}
+\frac{A_1^s}{\epsilon} +A_0^s)
\end{eqnarray}
with
\begin{eqnarray*}
&& A_2^s=\frac{13}{3},
\\
&& A_1^s= -2A_2^s\ln\delta_s +\frac{4}{3} +\frac{1}{3}
\ln\frac{-u_1}{m_{\tilde{t}_i}^2} -3
\ln\frac{-t_1}{m_{\tilde{t}_i}^2}
+\frac{4}{3}\ln\frac{s}{m_{\tilde{t}_i}^2},
\\
&& A_0^s=2 A_2^s\ln^2\delta_s -2A_1^s \ln\delta_s +
(C_F\frac{\gamma}{\beta} -\frac{1}{4}) \ln\frac{\gamma
+\beta}{\gamma -\beta} +\frac{1}{2}\ln^2\frac{s(\beta
-\gamma)}{t_1} + {\rm Li}_2[1 + \frac{t_1}{s(\gamma -\beta)}]
\\
&& \hspace{1.0cm} -{\rm Li}_2[1 +\frac{s(\gamma +\beta)}{t_1}]
+\frac{1}{6} \{\ln^2\frac{s(\beta - \gamma)}{u_1}
-\frac{1}{2}\ln^2\frac{\gamma +\beta}{\gamma -\beta} +2{\rm
Li}_2[\frac{t_1 +s (\beta +\gamma)}{s(\beta -\gamma)}]
\\
&& \hspace{1.0cm} -2{\rm Li}_2[\frac{t_1 +s
(\beta-\gamma)}{-u_1}]\},
\end{eqnarray*}
where $\gamma=(s + m_{\tilde{t}_i}^2
-m_{\tilde{\chi}_k^-}^2)/(2s)$ and
$\beta=\sqrt{\gamma^2-m_{\tilde{t}_i}^2/s}$.

\subsubsection{Hard collinear gluon emission}
In the hard collinear region, i.e. $E_3> \delta_s\sqrt{s}/2$ and
$0< -t' \ (-u') < \delta_c s$, the emitted hard gluon is collinear
to one of the incoming partons. As a consequence of the
factorization theorems \cite{factor1}, the squared matrix element
for $gb\rightarrow \tilde{t}_i \tilde{\chi}^-_k +g$ can be
factorized into the product of the Born squared matrix element and
the Altarelli-Parisi splitting function for $b\rightarrow bg$ and
$g\rightarrow gg$ \cite{altarelli1,factor2}, i.e.
\begin{eqnarray}
\overline{|M_{ik}^{R}(gb\rightarrow \tilde{t}_i \tilde{\chi}^-_k
+g)|}^2 \stackrel{collinear}{\rightarrow} (4\pi\alpha_s
\mu_r^{2\epsilon}) \overline{|M_{ik}^{B}|}^2
(\frac{-2P_{bb}(z,\epsilon)}{zt'}
+\frac{-2P_{gg}(z,\epsilon)}{zu'}),
\end{eqnarray}
where $z$ denotes the fraction of incoming parton $b(g)$'s
momentum carried by parton $b(g)$ with the emitted gluon taking a
fraction $(1-z)$, and $P_{ij}(z,\epsilon)$ are the unregulated
splitting functions in $n=4-2\epsilon$ dimensions for $0<z<1$,
which can be related to the usual Altarelli-Parisi splitting
kernels \cite{altarelli1} as $P_{ij}(z,\epsilon)=P_{ij}(z)
+\epsilon P_{ij}'(z)$, explicitly
\begin{eqnarray}
&& P_{qq}(z)=C_F \frac{1 +z^2}{1-z}, \hspace{4.5cm} P_{qq}'(z)=
-C_F (1-z), \nonumber \\
&& P_{gg}(z) =2 N [\frac{z}{ 1-z} +\frac{1-z}{z} +z(1-z)],
\hspace{1.0cm} P_{gg}'(z)=0.
\end{eqnarray}
Moreover, the three-body phase space also can be factorized in the
collinear limit, and, for example, in the limit $0< -t' < \delta_c
s$ it has the following form \cite{cutoff}:
\begin{eqnarray}
d\Gamma_3(gb\rightarrow \tilde{t}_i \tilde{\chi}^-_k +g)
\stackrel{collinear}{\rightarrow} d\Gamma_2(gb\rightarrow
\tilde{t}_i \tilde{\chi}^-_k; s'=zs)
\frac{(4\pi)^\epsilon}{16\pi^2\Gamma(1- \epsilon)} dz dt'[-(1
-z)t']^{-\epsilon}.
\end{eqnarray}
Note that the two-body phase space should be evaluated at a
squared parton-parton energy of $zs$. Then the three-body cross
section in the hard collinear region is given by \cite{cutoff}
\begin{eqnarray}
&& d\sigma^{HC}_{ik} =d\hat{\sigma}^B_{ik} [\frac{\alpha_s}{2\pi}
\frac{\Gamma(1-\epsilon)} {\Gamma(1-2\epsilon)}
(\frac{4\pi\mu_r^2}{s})^\epsilon] (-\frac{1}{\epsilon})
\delta_c^{-\epsilon} [P_{bb}(z,\epsilon)G_{b/p}(x_1/z)G_{g/p}(x_2)
\nonumber
\\ && \hspace{1.4cm} + P_{gg}
(z,\epsilon)G_{g/p}(x_1/z) G_{b/p}(x_2) +(x_1\leftrightarrow x_2)]
\frac{dz}{z} (\frac{1 -z}{z})^{-\epsilon} dx_1 dx_2,
\end{eqnarray}
where $G_{b(g)/p}(x)$ is the bare PDF.

\subsection{Final States with an Additional Massless Quark}
In addition to the real gluon emission, other real emission
corrections to the inclusive cross section for $pp\rightarrow
\tilde{t}_i \tilde{\chi}^-_k$ at NLO involve the processes with an
additional massless (anti)quark in the final states, as shown in
Fig.5 ($ q=u,d,s,c$):
\begin{eqnarray}
&& g(p_a)+ g(p_b) \rightarrow \tilde{t}_i(p_1)+ \tilde{\chi}_k^-
(p_2) + \bar{b}(p_3), \label{gg} \\
&& q/\bar{q}(p_a) + b(p_b) \rightarrow \tilde{t}_i(p_1) +
\tilde{\chi}_k^- (p_2) + q/\bar{q}(p_3), \label{qb} \\
&& b/\bar{b}(p_b) +b(p_a) \rightarrow \tilde{t}_i (p_1) +
\tilde{\chi}_k^- (p_2) + b/\bar{b}(p_3), \label{bb} \\
&& q(p_a) + \bar{q}(p_b) \rightarrow \tilde{t}_i(p_1) +
\tilde{\chi}_k^-(p_2) + \bar{b}(p_3) \label{qbq}.
\end{eqnarray}
Since the contributions from the processes (\ref{gg})-(\ref{bb})
contain the initial state collinear singularities, we also need to
use the two cutoff phase space slicing method \cite{cutoff}.
However, we only split the phase space into two regions, because
there are no soft divergences here. Thus, according to the
approach shown in Ref.~\cite{cutoff}, the cross sections for the
processes with an additional massless (anti)quark in the final
states can be expressed as ($ q=u,d,s,c,b$)
\begin{eqnarray}
&& d\sigma_{ik}^{add}= \sum_{(\alpha,\beta)}
\hat{\sigma}_{ik}^{\overline{C}}(\alpha\beta\rightarrow
\tilde{t}_i \tilde{\chi}_k^- +X) [G_{\alpha/p}(x_1)
G_{\beta/p}(x_2) +(x_1\leftrightarrow x_2)] dx_1dx_2 \nonumber
\\&& \hspace{1.0cm}
+d\hat{\sigma}^B_{ik} [\frac{\alpha_s}{2\pi}
\frac{\Gamma(1-\epsilon)} {\Gamma(1-2\epsilon)}
(\frac{4\pi\mu^2_r}{s})^\epsilon] (-\frac{1}{\epsilon})
\delta_c^{-\epsilon} [P_{bg}(z,\epsilon)G_{g/p}(x_1/z)G_{g/p}(x_2)
\nonumber
\\ && \hspace{1.0cm} + \sum_{\alpha=q,\bar{q}} P_{g\alpha}
(z,\epsilon)G_{\alpha/p}(x_1/z) G_{b/p}(x_2) +(x_1\leftrightarrow
x_2)] \frac{dz}{z} (\frac{1 -z}{z})^{-\epsilon} dx_1 dx_2,
\label{add}
\end{eqnarray}
where
\begin{eqnarray}
&& P_{gq}(z) =C_F \frac{1 +(1-z)^2}{z}, \hspace{2.0cm} P_{gq}'(z)=
-C_F z, \nonumber
\\ &&
P_{qg}(z) =\frac{1}{2}[z^2 +(1-z)^2], \hspace{2.0cm}
P_{qg}'(z)=-z(1-z).
\end{eqnarray}
The first term in Eq.(\ref{add}) represents the non-collinear
cross sections for the four processes, and like Eq.(\ref{nonHC})
each $\hat{\sigma}_{ik}^{\overline{C}}$ can be written in the
form:
\begin{eqnarray}
d\hat{\sigma}^{\overline{C}}_{ik}=\frac{1}{2s}
\overline{|M_{ik}^{\alpha\beta}|}^2 d\overline{\Gamma}_3,
\label{qHC}
\end{eqnarray}
where $\alpha$ and $\beta$ denote the incoming partons in the
partonic processes, and $d\overline{\Gamma}_3$ is the three body
phase space in the non-collinear region. The explicit expressions
of $\overline{|M_{ik}^{\alpha\beta}|}^2$ can be found in Appendix
B. The second term in Eq.(\ref{add}) represents the collinear
singular cross sections.

Moreover, for the subprocesses $gg/q\bar{q} \rightarrow
\tilde{t}_i\bar{\tilde{t}}_i^\ast \rightarrow
\tilde{t}_i\tilde{\chi}_k^-\bar{b}$ ($q=u,d,s,c,b$), assuming
$m_{\tilde{t}_i} >m_{\tilde{\chi}_k^-}$, the stop momentum can
approach the stop mass shell, which will lead to singularity
arising from the stop propagator. Following the analysis shown in
Ref.~\cite{beenakker2}, this problem can easily be solved by
introducing the non-zero stop widths $\Gamma_{\tilde{t}_i}$ and
regularizing in this way the higher-order amplitudes. However,
these on-shell stop contributions are already accounted for by the
LO stop pair production with a subsequent decay into a chargino
and a bottom quark, and thus should not be considered as a genuine
higher order correction to the associated production of top
squarks and charginos. Therefore, to avoid double counting, these
pole contributions will be subtracted in our numerical
calculations below in the same way as shown in Appendix B of
Ref.~\cite{beenakker2}.

\subsection{Mass factorization}
As mentioned above, after adding the renormalized virtual
corrections and the real corrections, the partonic cross sections
still contain the collinear divergences, which can be absorbed
into the redefinition of the PDF at NLO, in general called mass
factorization \cite{altarelli}. This procedure in practice means
that first we convolute the partonic cross section with the bare
PDF $G_{\alpha/p}(x)$, and then use the renormalized PDF
$G_{\alpha/p}(x,\mu_f)$ to replace $G_{\alpha/p}(x)$. In the
$\overline{\rm MS}$ convention, the scale dependent PDF
$G_{\alpha/p}(x,\mu_f)$ is given by \cite{cutoff}
\begin{eqnarray}
G_{\alpha/p}(x,\mu_f)= G_{\alpha/p}(x)+
\sum_{\beta}(-\frac{1}{\epsilon}) [\frac{\alpha_s}{2\pi}
\frac{\Gamma(1 -\epsilon)}{\Gamma(1 -2\epsilon)} (\frac{4\pi
\mu_r^2}{\mu_f^2})^\epsilon] \int_x^1 \frac{dz}{z} P_{\alpha\beta}
(z) G_{\beta/p}(x/z).
\end{eqnarray}
This replacement will produce a collinear singular counterterm,
which is combined with the hard collinear contributions to result
in, as the definition in Ref.~\cite{cutoff}, the ${\cal O}
(\alpha_s)$ expression for the remaining collinear contribution:
\begin{eqnarray}
&& d\sigma_{ik}^{coll}=d\hat{\sigma}^B_{ik}[\frac{\alpha_s}{2\pi}
\frac{\Gamma(1-\epsilon)} {\Gamma(1-2\epsilon)}
(\frac{4\pi\mu^2_r}{s})^\epsilon] \{\tilde{G}_{g/p}(x_1,\mu_f)
G_{b/p}(x_2,\mu_f) + G_{g/p}(x_1,\mu_f) \tilde{G}_{b/p}(x_2,\mu_f)
\nonumber
\\ && \hspace{1.2cm}
+\sum_{\alpha=b,g}[\frac{A_1^{sc}(\alpha\rightarrow \alpha
g)}{\epsilon} +A_0^{sc}(\alpha\rightarrow \alpha
g)]G_{g/p}(x_1,\mu_f) G_{b/p}(x_2,\mu_f)
\nonumber
\\ && \hspace{1.2cm}
+(x_1\leftrightarrow x_2)\} dx_1dx_2,
\end{eqnarray}
where
\begin{eqnarray}
&& A_1^{sc}(q\rightarrow qg)=C_F(2\ln\delta_s +3/2), \\
&& A_1^{sc}(g\rightarrow gg)=2N\ln\delta_s +(11N -2n_f)/6, \\
&& A_0^{sc}=A_1^{sc}\ln(\frac{s}{\mu_f^2}), \\
&& \tilde{G}_{\alpha/p}(x,\mu_f)=\sum_{\beta}\int_x^{1-
\delta_s\delta_{\alpha\beta}} \frac{dy}{y}
G_{\beta/p}(x/y,\mu_f)\tilde{P}_{\alpha\beta}(y)
\end{eqnarray}
with
\begin{eqnarray}
\tilde{P}_{\alpha\beta}(y)=P_{\alpha\beta} \ln(\delta_c
\frac{1-y}{y} \frac{s}{\mu_f^2}) -P_{\alpha\beta}'(y).
\end{eqnarray}

Finally, the NLO total cross section for $pp\rightarrow
\tilde{t}_i\tilde{\chi}_k^-$ in the $\overline{MS}$ factorization
scheme is
\begin{eqnarray}
&& \sigma^{NLO}_{ik}= \int \{dx_1dx_2
[G_{b/p}(x_1,\mu_f)G_{g/p}(x_2,\mu_f)+ x_1\leftrightarrow
x_2](\hat{\sigma}^{B}_{ik} + \hat{\sigma}^{V}_{ik}+
\hat{\sigma}^{S}_{ik} +\hat{\sigma}^{\overline{HC}}_{ik})
+d\sigma_{ik}^{coll}\} \nonumber
\\ && \hspace{1.0cm} +\sum_{(\alpha,\beta)}\int dx_1dx_2
[G_{\alpha/p}(x_1,\mu_f) G_{\beta/p}(x_2,\mu_f)
+(x_1\leftrightarrow x_2)]
\hat{\sigma}_{ik}^{\overline{C}}(\alpha\beta\rightarrow
\tilde{t}_i \tilde{\chi}_k^- +X).
\end{eqnarray}
Note that the above expression contains no singularities since
$A_2^V +A_2^s =0$ and $A_1^V +A_1^s +A_1^{sc}(b\rightarrow bg)
+A_1^{sc}(g\rightarrow gg) =0$.

\subsection{Differential Cross section in the Transverse Momentum}
In this subsection we present the differential cross section in
the transverse momentum $p_T$ of the charginos. Using the
notations defined in Ref.~\cite{beenakker2}, the differential
distribution with respect to $p_T$ and $y$ for the processes
\begin{eqnarray}
p(P_a)+ p(P_b) \rightarrow \tilde{t}_i (p_1) +\tilde{\chi}_k^-
(p_2) [+ g(p_3)/q(p_3)/\bar{q}(p_3)]
\end{eqnarray}
is given by
\begin{eqnarray} \label{integralpt}
\frac{d^2\sigma}{dp_T dy} =2 p_T S \sum_{\alpha,\beta}
\int_{x_1^-}^1 dx_1 \int_{x_2^-}^1 dx_2 x_1
G_{\alpha/p}(x_1,\mu_f) x_2 G_{\beta/p}(x_2,\mu_f) \frac{d^2
\hat{\sigma}_{\alpha\beta}}{dt_2 du_2},
\end{eqnarray}
where $\sqrt{S}$ is the total center-of-mass energy of the
collider, and
\begin{eqnarray}
&& p_T^2= \frac{T_2U_2}{S} -m_{\tilde{\chi}_k^-}^2, \hspace{1.8cm}
y=\frac{1}{2}\ln(\frac{T_2}{U_2}), \nonumber \\ && x_1^-=
\frac{-T_2 -m_{\tilde{\chi}_k^-}^2 +m_{\tilde{t}_i}^2}{S +U_2}, \
\ \ \ \ \ x_2^-= \frac{-x_1U_2 -m_{\tilde{\chi}_k^-}^2
+m_{\tilde{t}_i}^2}{x_1S +T_2}
\end{eqnarray}
with $T_2=(P_b-p_2)^2 - m_{\tilde{\chi}_k^-}^2$ and
$U_2=(P_a-p_2)^2 - m_{\tilde{\chi}_k^-}^2$. The limits of integral
over $y$ and $p_T$ are
\begin{eqnarray}
-y^{max}(p_T)\leq y \leq y^{max}(p_T), \hspace{1.5cm}  0\leq p_T
\leq p_T^{max},
\end{eqnarray}
with
\begin{eqnarray}
&& y^{max}(p_T)={\rm arccosh}(\frac{S+ m_{\tilde{\chi}_k^-}^2-
m_{\tilde{t}_i}^2}{2\sqrt{S (p_T^2 +m_{\tilde{\chi}_k^-}^2)}}),
\nonumber
\\ && p_T^{max}=\frac{1}{2\sqrt{S}} \sqrt{(S
+m_{\tilde{\chi}_k^-}^2 -m_{\tilde{t}_i}^2)^2
-4m_{\tilde{\chi}_k^-}^2S}.
\end{eqnarray}

\section{Numerical Results and Conclusion}
We now present the numerical results for total and differential
cross sections for the associated production of top squarks and
charginos at the LHC. In our numerical calculations the Standard
Model (SM) parameters were taken to be $\alpha_{ew}(m_W)=1/128$,
$m_W=80.419$ GeV, $m_Z=91.1882$ GeV, and $m_t=174.3$ GeV
\cite{SM}. We use the two-loop evolution of $\alpha_s(Q)$
\cite{runningalphas} ($\alpha_s(M_Z)=0.118$) and CTEQ6M (CTEQ6L)
PDFs \cite{CTEQ} throughout the calculations of the NLO (LO) cross
sections. Moreover, in order to improve the perturbative
calculations, we take the running mass $m_b(Q)$ evaluated by the
NLO formula \cite{runningmb}:
\begin{equation}
m_b(Q)=U_6(Q,m_t)U_5(m_t,m_b)m_b(m_b)
\end{equation}
with $m_b(m_b)=4.25$ GeV \cite{mb}. The evolution factor $U_f$ is
\begin{eqnarray}
U_f(Q_2,Q_1)=(\frac{\alpha_s(Q_2)}{\alpha_s(Q_1)})^{d^{(f)}}
[1+\frac{\alpha_s(Q_1)-\alpha_s(Q_2)}{4\pi}J^{(f)}], \nonumber \\
d^{(f)}=\frac{12}{33-2f}, \hspace{1.0cm}
J^{(f)}=-\frac{8982-504f+40f^2}{3(33-2f)^2}.
\end{eqnarray}
In addition, in order to improve the perturbation calculations,
especially for large $\tan\beta$, we make the following
replacement in the tree-level couplings \cite{runningmb}:
\begin{eqnarray}
&& m_b(Q) \ \ \rightarrow \ \ \frac{m_b(Q)}{1+\Delta m_b},
\label{deltamb}
\\
&& \Delta m_b=\frac{2\alpha_s}{3\pi}M_{\tilde{g}}\mu\tan\beta
I(m_{\tilde{b}_1},m_{\tilde{b}_2},M_{\tilde{g}})
+\frac{h_t^2}{16\pi^2}\mu A_t\tan\beta
I(m_{\tilde{t}_1},m_{\tilde{t}_2},\mu) \nonumber \\
&& \hspace{1.0cm} -\frac{g^2}{16\pi^2}\mu M_2\tan\beta
\sum_{i=1}^2 [(R^{\tilde{t}}_{i1})^2 I(m_{\tilde{t}_i},M_2,\mu) +
\frac{1}{2}(R^{\tilde{b}}_{i1})^2 I(m_{\tilde{b}_i},M_2,\mu)]
\label{deltamb1}
\end{eqnarray}
with
\begin{eqnarray}
I(a,b,c)=\frac{1}{(a^2-b^2)(b^2-c^2)(a^2-c^2)}
(a^2b^2\log\frac{a^2}{b^2} +b^2c^2\log\frac{b^2}{c^2}
+c^2a^2\log\frac{c^2}{a^2}).
\end{eqnarray}
And, it is necessary for avoiding double counting to subtract
these (SUSY-)QCD corrections from the renormalization constant
$\delta m_b$ in the following numerical calculations. The MSSM
parameters are determined as follows:

(i) For the parameters $M_1$, $M_2$ and $\mu$ in the chargino and
neutralino matrices, we take $M_2$ and $\mu$ as the input
parameters, and assuming gaugino mass unification we take
$M_1=(5/3)\tan^2\theta_W M_2$ and $M_{\tilde{g}}=
(\alpha_s(M_{\tilde{g}})/\alpha_2)M_2$ \cite{Hidaka}.

(ii) For the parameters in squark mass matrices, we assume
$M_{\tilde Q}=M_{\tilde U}=M_{\tilde D}$ and $A_t=A_b=300$ GeV to
simplify the calculations. Actually, the numerical results are not
very sensitive to $A_{t(b)}$.

In our calculations, except in Fig.10, we always choose the
renormalization scale $\mu_r=m_{\rm av}$, and the factorization
scale $\mu_f$ is fixed to $m_{\rm av}/3$ instead of $m_{\rm av}$
as mentioned in Sec. I.

Before we discuss in detail the numerical results for associated
production of top squarks and charginos at the LHC, in Fig.6 we
first show that the following NLO QCD predictions do not depend on
the arbitrary cutoffs $\delta_s$ and $\delta_c$ in the two cutoff
phase space slicing method. Here we take $\mu=-200$ GeV, $M_2=300$
GeV, $m_{\tilde{t}_1}=250$ GeV, $\tan\beta=30$, and
$\delta_c=\delta_s/100$. $\sigma_{other}$ includes the
contributions from the Born cross section and the virtual
corrections, which are cutoff-independent. We can see that the
soft plus hard collinear contributions and the hard non-collinear
contributions depend strongly on the cutoffs, and especially for
the small cutoffs ($\delta_s<10^{-4}$) their magnitudes can be ten
times larger than the LO total cross section. However, the two
contributions ($\sigma_{soft}+\sigma_{hard/coll}$ and
$\sigma_{hard/non-coll}$) balance each other very well, and the
final result $\sigma_{NLO}$ is independent of the cutoffs and
keeps 0.23 pb. Therefore, we take $\delta_s=10^{-4}$ and
$\delta_c=\delta_s/100$ in the numerical calculations of
Fig.7--13.

Fig.7 shows the dependence of the LO and NLO predictions for
$pp\rightarrow \tilde{t}_i\tilde{\chi}_k^-$ on $m_{\tilde{t}_1}$,
assuming $\mu=-200$ GeV, $M_2=300$ GeV and $\tan\beta=30$, which
means that two chargino masses are about $182$ GeV and $331$ GeV,
respectively, and $m_{\tilde{t}_2}$ increases from 342 GeV to 683
GeV for $m_{\tilde{t}_1}$ in the range 100--600 GeV. One finds
that the total cross sections for the
$\tilde{t}_2\tilde{\chi}_2^-$ channel are always smallest and less
than 3 fb, but the total cross sections for other channels are
large and range between $10$ fb and several hundred fb for most
values of $m_{\tilde{t}_1}$. Especially for the
$\tilde{t}_1\tilde{\chi}_1^-$ channel, the total cross section can
reach 1 pb for small values of $m_{\tilde{t}_1}$ (100 GeV $<
m_{\tilde{t}_1}<$ 160 GeV), which is almost the same as ones of
top quark and charged Higgs boson associated production at the
LHC. However, when $m_{\tilde{t}_1}$ get larger and close to
$m_{\tilde{t}_2}$, the total cross section for the
$\tilde{t}_2\tilde{\chi}_1^-$ channel is the largest, as shown in
Fig.7. Moreover, Fig.7 also shows that the NLO QCD corrections
enhance the LO results significantly, which are in general a few
ten percent.

In Fig.8 we show the dependence of the LO and NLO predictions for
$pp\rightarrow \tilde{t}_i\tilde{\chi}_k^-$ on
$m_{\tilde{\chi}_1^-}$, assuming $\mu=-400$ GeV,
$m_{\tilde{t}_1}=250$ GeV and $\tan\beta=30$, which means that
$m_{\tilde{t}_2}=414$ GeV and $m_{\tilde{\chi}_2^-}$ increases
from 417 GeV to 434 GeV for $m_{\tilde{\chi}_1^-}$ in the range
100--300 GeV. One can see that the total cross sections for all
channels can exceed $10$ fb. For small values of
$m_{\tilde{\chi}_1^-}$, the $\tilde{t}_1\tilde{\chi}_1^-$ channel
has the largest total cross sections, and for large values of
$m_{\tilde{\chi}_1^-}$, the $\tilde{t}_1\tilde{\chi}_2^-$ channel
has the largest ones, which all can exceed 100 fb. Moreover, Fig.8
also shows that the NLO QCD corrections enhance the LO results,
and especially for the $\tilde{t}_1\tilde{\chi}_1^-$ and
$\tilde{t}_2\tilde{\chi}_1^-$ channels, the enhancement can exceed
$50 \%$.

The associated production of $\tilde{t}_1$ and $\tilde{\chi}_1^-$
is the most important since the total cross sections are the
largest for $m_{\tilde{t}_1}<$ 400 GeV and $m_{\tilde{\chi}_1^-}<$
230 GeV. Thus we mainly discuss this channel below.

Fig.9 gives the dependence of the $K$ factor (the ratio of the NLO
total cross section over the LO one) on $m_{\tilde{t}_1}$ for the
$\tilde{t}_1\tilde{\chi}_1^-$ production. Here we assume the same
values of the MSSM parameters as in Fig.7. The $K$ factor contains
four contributions: (i) The curve $(b)$ shows the ratio of the
improved Born cross section, which is defined as convoluting the
LO partonic cross section with the NLO PDFs, over the LO one. The
corresponding $K$ factor ranges from 1.18 to 1.39. (ii) The NLO
cross section in the curve $(c)$ contains only the contributions
from the subprocess with two initial-state gluons, which are
positive and the corresponding $K$ factor varies from $0.30$ to
$0.06$. (iii) The NLO cross section in the curve $(d)$ collects
only the contributions from massless (anti)quark emission
subprocesses as shown in Eq.(\ref{qb})-(\ref{qbq}), and the
corresponding $K$ factor keeps always about $0.10$. (iv) The NLO
cross section in the curve $(e)$ includes only the virtual and the
real gluon emission corrections, which are negative and the
corresponding $K$ factor ranges between $-0.36$ and $-0.29$.
Finally, summing above four parts leads to the total $K$ factor as
shown in the curve $(a)$, which indicates that the NLO QCD
corrections enhance the LO total cross section and vary from
$21\%$ to $27\%$.

In Fig.10 we show the dependence of the total cross sections for
the $\tilde{t}_1\tilde{\chi}_1^-$ production on the
renormalization/factorization scale, assuming $\mu=-200$ GeV,
$M_2=300$ GeV, $\tan\beta=30$, $m_{\tilde{t}_1}=250$ GeV, and
$\mu_r=\mu_f$. One finds that the NLO QCD corrections reduce the
dependence significantly. The cross sections vary by $\pm 15\%$ at
LO but only by $\pm 4\%$ at NLO in the region $0.5<\mu_f/m_{\rm
av}<2.0$. Thus the reliability of the NLO predictions has been
improved substantially.

The cross sections of the $\tilde{t}_1\tilde{\chi}_1^-$ production
as a function of $\tan\beta$ are displayed for
$m_{\tilde{t}_1}=200, 300$ and 400 GeV in Fig.11, assuming
$\mu=-200$ GeV and $M_2=300$ GeV. For the dotted curves, $\Delta
m_b$ in Eq.(\ref{deltamb}) is replaced by $\Delta m^\prime_b$
which only includes the corrections of the gluino piece and
neglects the rest of the SUSY diagrams in Eq.(\ref{deltamb1}).
From Fig.11, one can see that the cross sections become large when
$\tan\beta$ gets high or low, which is due to the fact that for
the coupling $b-\tilde{t}_1-\tilde{\chi}_1^-$ at low $\tan\beta$
the top quark contribution is enhanced while at high $\tan\beta$
the bottom quark contribution becomes large. Fig.11 also shows
that the NLO QCD corrections enhance the LO total cross sections,
and for $m_{\tilde{t}_1}=200$ and $300$ GeV, the enhancement is
more significant for the medium values of $\tan\beta$ than for the
other values. Moreover, after comparing the solid curves with the
dotted ones, we can see that the SUSY corrections to $\Delta m_b$
except the gluino piece decrease the total cross sections
significantly, especially for larger values of $\tan\beta$.

Fig.12 shows the dependence of the total cross sections for the
$\tilde{t}_1\tilde{\chi}_1^-$ production on the parameters $\mu$
and $M_2$, assuming $m_{\tilde{t}_1}=250$ GeV and $\tan\beta=30$.
Note that the phenomenology at CERN $e^+e^-$ LEP and Tevatron
\cite{exclude} has excluded the parameter range $|\mu|\lesssim
180$ GeV, where one chargino mass and one of the neutralino masses
can become very small, and especially the latter is zero for
vanishing $\mu$. So we focus on the range $|\mu|\gtrsim 180$ GeV.
One can see that the cross sections increase with an increase of
$M_2$ for $|\mu|<450$ GeV, and decrease with an increase of
$|\mu|$ except $|\mu|>350$ GeV for $M_2=300$ GeV. The NLO QCD
corrections can increase and decrease the LO total cross sections
depending on the values of $\mu$ and $M_2$.

In Fig.13 we display the differential cross sections in the
transverse momentum $p_T$ of $\tilde{\chi}_1^-$ for the
$\tilde{t}_1\tilde{\chi}_1^-$ production in three cases, which are
obtained after integration over all $y$ in Eq.(\ref{integralpt}),
assuming $\mu=-200$ GeV and $m_{\tilde{t}_1}=250$ GeV. In all
cases, we find that the NLO QCD corrections enhance the LO
differential cross sections for low $p_T$, but decrease the LO
results for high $p_T$.

In conclusion, we have calculated the NLO inclusive total cross
sections for the associated production processes $pp\rightarrow
\tilde{t}_i\tilde{\chi}_k^-$ in the MSSM at the LHC. Our
calculations show that the total cross sections for the
$\tilde{t}_1\tilde{\chi}_1^-$ production for the lighter top
squark masses in the region 100 GeV $< m_{\tilde{t}_1}<$ 160 GeV
can reach 1 pb in the favorable parameter space allowed by the
current precise experiments, and in other cases the total cross
sections generally vary from $10$ fb to several hundred fb except
both $m_{\tilde{t}_1}>$ 500 GeV and the
$\tilde{t}_2\tilde{\chi}_2^-$ production channel, which means that
the LHC may produce abundant events of these processes, and it is
very possible to discover these SUSY particles through the above
processes in the future experiments, if the supersymmetry exists.
Moreover, we find that the NLO QCD corrections in general enhance
the LO total cross sections significantly, and vastly reduce the
dependence of the total cross sections on the
renormalization/factorization scale, which leads to increased
confidence in predictions based on these results.

\begin{acknowledgments}
We would like to thank T. Plehn, C.-P. Yuan, T.M.P. Tait and M.
Klasen for useful discussions and valuable suggestions. This work
was supported in part by the National Natural Science Foundation
of China.
\end{acknowledgments}
\appendix
%%%%%%%%%%%%%%%%%%%%%%%%%%%% appendix A %%%%%%%%%%%%%%%%%%%%%%%
\section{}
In this appendix, we collect the explicit expressions of the
nonzero form factors in Eq.(\ref{fvertex}) and Eq.(\ref{fbox}).
For simplicity, we introduce the following abbreviations for the
Passarino-Veltman three-point integrals $C_{i(j)}$ and four-point
integrals $D_{i(j)}$, which are defined similar to
Ref.~\cite{denner} except that we take internal masses squared as
arguments:
\begin{eqnarray*}
&& C_{i(j)}^a= C_{i(j)}(0, 0, s; 0,0,0),
\\
&& C_{i(j)}^b= C_{i(j)}(m_{\tilde{t}_i}^2, m_{\tilde{\chi}_k^-}^2,
s; 0,m_{\tilde{t}_i}^2,0),
\\
&& C_{i(j)}^c= C_{i(j)}(t, m_{\tilde{t}_i}^2, 0;m_{\tilde{t}_i}^2,
0,m_{\tilde{t}_i}^2),
\\
&& C_{i(j)}^d= C_{i(j)}(0, m_{\tilde{t}_i}^2, t; 0,
0,m_{\tilde{t}_i}^2),
\\
&& C_{i(j)}^e= C_{i(j)}(m_{\tilde{\chi}_k^-}^2, 0, t;
m_{\tilde{t}_i}^2, 0, 0),
\\
&& C_{i(j)}^f= C_{i(j)}(0, 0, s; m_{\tilde{g}}^2, m_{\tilde{g}}^2,
m_{\tilde{b}_j}^2),
\\
&& C_{i(j)}^g= C_{i(j)}(s, 0, 0; m_{\tilde{b}_j}^2,
m_{\tilde{g}}^2, m_{\tilde{b}_j}^2),
\\
&& C_{i(j)}^h= C_{i(j)}(t, m_{\tilde{t}_i}^2, 0; m_{\tilde{g}}^2,
m_t^2, m_{\tilde{g}}^2),
\\
&& C_{i(j)}^i= C_{i(j)}(t, m_{\tilde{\chi}_k^-}^2, 0;
m_{\tilde{g}}^2, m_t^2, m_{\tilde{b}_j}^2),
\\
&& C_{i(j)}^j= C_{i(j)}(m_{\tilde{t}_i}^2, m_{\tilde{\chi}_k^-}^2,
s; m_{\tilde{g}}^2, m_t^2, m_{\tilde{b}_j}^2),
\\
&& C_{i(j)}^k= C_{i(j)}(m_{\tilde{t}_i}^2,0,u; m_t^2,
m_{\tilde{g}}^2, m_{\tilde{b}_j}^2),
\\
&& D_{i(j)}^a = D_{i(j)}(m_{\tilde{t}_i}^2,
t, 0, s, 0, m_{\tilde{\chi}_k^-}^2; 0, m_{\tilde{t}_i}^2, 0, 0),
\\
&& D_{i(j)}^b = D_{i(j)}(m_{\tilde{t}_i}^2, u, 0, s, 0,
m_{\tilde{\chi}_k^-}^2; 0, m_{\tilde{t}_i}^2, 0, 0),
\\
&& D_{i(j)}^c = D_{i(j)}(t, m_{\tilde{\chi}_k^-}^2, u,
m_{\tilde{t}_i}^2, 0, 0; 0, m_{\tilde{t}_i}^2, 0,
m_{\tilde{t}_i}^2),
\\
&& D_{i(j)}^d = D_{i(j)}(0, t, m_{\tilde{\chi}_k^-}^2, s,
m_{\tilde{t}_i}^2, 0; m_{\tilde{g}}^2, m_{\tilde{g}}^2, m_t^2,
m_{\tilde{b}_j}^2),
\\
&& D_{i(j)}^e = D_{i(j)}(s, m_{\tilde{\chi}_k^-}^2, u, 0,
m_{\tilde{t}_i}^2, 0; m_{\tilde{g}}^2, m_{\tilde{b}_j}^2, m_t^2,
m_{\tilde{b}_j}^2),
\\
&& D_{i(j)}^f = D_{i(j)}(t, m_{\tilde{\chi}_k^-}^2, u,
m_{\tilde{t}_i}^2, 0, 0; m_{\tilde{g}}^2, m_t^2,
m_{\tilde{b}_j}^2, m_t^2).
\end{eqnarray*}
Many above functions contain the soft infrared and collinear
singularities, but all Passarino-Veltman integrals can be reduced
down to the scalar function $B_0$, $C_0$ and $D_0$. Here we
present the explicit expressions of $C_0$ and $D_0$, which contain
the singularities and are used in our calculations:
\begin{eqnarray*}
&& C_0^a=\frac{C_{\epsilon}} {s}(\frac{1}{\epsilon^2}
-\frac{1}{\epsilon}\ln\frac{s}{m_{\tilde{t}_i}^2}
+\frac{1}{2}\ln^2\frac{s}{m_{\tilde{t}_i}^2} -\frac{2\pi^2}{3}),
\\
&& C_0^d=\frac{C_{\epsilon}} {t_1} [\frac{1}{2\epsilon^2}
-\frac{1}{\epsilon} \ln(\frac{-t_1}{m_{\tilde{t}_i}^2}) -{\rm
Li}_2(\frac{t}{t_1}) + \frac{1}{2}
\ln^2(\frac{-t_1}{m_{\tilde{t}_i}^2})],
\\
&& C_0^e=\frac{C_{\epsilon}} {t_2} [\frac{1}{\epsilon}
\ln(\frac{m_{\tilde{\chi}_k^-}^2 -m_{\tilde{t}_i}^2}{t_1}) +{\rm
Li}_2(\frac{m_{\tilde{\chi}_k^-}^2} {m_{\tilde{\chi}_k^-}^2
-m_{\tilde{t}_i}^2}) -{\rm Li}_2(\frac{t}{t_1})
-\frac{1}{2}\ln^2(1-
\frac{m_{\tilde{\chi}_k^-}^2}{m_{\tilde{t}_i}^2})
+\frac{1}{2}\ln^2(\frac{-t_1}{m_{\tilde{t}_i}^2})],
\\
&& D_0^a=\frac{C_{\epsilon}} {st_1}[\frac{3}{2\epsilon^2}
-\frac{1}{\epsilon} \ln\frac{s}{m_{\tilde{t}_i}^2}
+\frac{1}{\epsilon} \ln\frac{m_{\tilde{t}_i}^2 -
m_{\tilde{\chi}_k^-}^2} {m_{\tilde{t}_i}^2} -\frac{2}{\epsilon}
\ln\frac{-t_1} {m_{\tilde{t}_i}^2} -\frac{\pi^2}{2} -{\rm
Li}_2(\frac{t}{t_1}) +{\rm Li}_2(\frac{t_2}{m_{\tilde{t}_i}^2 -
m_{\tilde{\chi}_k^-}^2})
\\
&& \hspace{1.0cm} -{\rm Li}_2(1 +\frac{m_{\tilde{t}_i}^2 -
m_{\tilde{\chi}_k^-}^2}{s}) +2{\rm Li}_2(-\frac{f_b}{st_1})
-\frac{1}{2}\ln^2\frac{m_{\tilde{t}_i}^2 - m_{\tilde{\chi}_k^-}^2}
{m_{\tilde{t}_i}^2} +\ln\frac{s}{m_{\tilde{t}_i}^2}
\ln\frac{m_{\tilde{t}_i}^2 - m_{\tilde{\chi}_k^-}^2}
{m_{\tilde{t}_i}^2}
\\
&& \hspace{1.0cm} +\frac{1}{2}\ln^2\frac{- t_1}
{m_{\tilde{t}_i}^2} +\ln\frac{st_1+f_b}{st_1}
(\ln\frac{m_{\tilde{\chi}_k^-}^2 -m_{\tilde{t}_i}^2}{s}-
\ln\frac{- t_1} {m_{\tilde{t}_i}^2}) -{\rm R}(0,1, -t_1,t_2,0,
-\frac{s}{m_{\tilde{t}_i}^2})
\\
&& \hspace{1.0cm} -{\rm R}(0,1,st_1,f_b,-1,-\frac{t}{t_1}) -{\rm
R}(0,1,st_1,f_b,-1,1+\frac{m_{\tilde{t}_i}^2
-m_{\tilde{\chi}_k^-}^2}{s})],
\\
&& D_0^b=D_0^a \ (t\rightarrow u, \ t_1 \rightarrow u_1, \ t_2
\rightarrow u_2),
\\
&& D_0^c=\frac{C_{\epsilon}} {u_1t_1}[\frac{1}{2\epsilon^2}
-\frac{1}{\epsilon} \ln\frac{-t_1}{m_{\tilde{t}_i}^2}
+\frac{1}{\epsilon} \ln\frac{m_{\tilde{t}_i}^2 -
m_{\tilde{\chi}_k^-}^2} {m_{\tilde{t}_i}^2} -\frac{1}{\epsilon}
\ln\frac{-u_1} {m_{\tilde{t}_i}^2} -{\rm Li}_2(\frac{t}{t_1})
+{\rm Li}_2(\frac{m_{\tilde{\chi}_k^-}^2}{m_{\tilde{\chi}_k^-}^2 -
m_{\tilde{t}_i}^2})
\\
&& \hspace{1.0cm} -{\rm Li}_2(\frac{u}{u_1})
-\frac{1}{2}\ln^2\frac{m_{\tilde{t}_i}^2 - m_{\tilde{\chi}_k^-}^2}
{m_{\tilde{t}_i}^2}
+\frac{1}{2}\ln^2\frac{-t_1}{m_{\tilde{t}_i}^2}
+\frac{1}{2}\ln^2\frac{- u_1} {m_{\tilde{t}_i}^2} +{\rm
R}(\frac{u}{m_{\tilde{\chi}_k^-}^2},1,\frac{-f_c}{m_{\tilde{\chi}_k^-}^2},s,1,-
\frac{m_{\tilde{\chi}_k^-}^2}{m_{\tilde{t}_i}^2})
\\
&& \hspace{1.0cm} +{\rm
R}(\frac{t}{m_{\tilde{t}_i}^2},1,\frac{-f_c}{m_{\tilde{t}_i}^2},u,1,-1)
+{\rm R}(0,1, u_1t_1,f_c,\frac{m_{\tilde{t}_i}^2
-m_{\tilde{\chi}_k^-}^2}{m_{\tilde{t}_i}^2},
\frac{m_{\tilde{\chi}_k^-}^2}{m_{\tilde{t}_i}^2})
\\
&& \hspace{1.0cm} -{\rm R}(0,1,u_1t_1,f_c,-\frac{u_1}
{m_{\tilde{t}_i}^2}, \frac{u}{m_{\tilde{t}_i}^2}) -{\rm
R}(0,1,u_1t_1,f_c,-\frac{t_1}{m_{\tilde{t}_i}^2},
\frac{t}{m_{\tilde{t}_i}^2})],
\end{eqnarray*}
where we define $C_{\epsilon}=\Gamma(1 +\epsilon) (4\pi
\mu_r^2/m_{\tilde{t}_i}^2)^{\epsilon}$, and
\begin{eqnarray*}
&& f_b=t(t_1+u_1) +m_{\tilde{t}_i}^2(m_{\tilde{t}_i}^2
-m_{\tilde{\chi}_k^-}^2), \hspace{1.0cm}
f_c=m_{\tilde{t}_i}^2m_{\tilde{\chi}_k^-}^2 -tu,
\\
&& {\rm R}(x_1,x_2,x_3,x_4,x_5,x_6) ={\rm Li}_2(\frac{x_6 x_1'}
{\Delta'}) -{\rm Li}_2(\frac{x_6 x_2'} {\Delta'})
+\ln\frac{-\Delta'}{x_4} \ln\frac{x_2'}{x_1'}
\end{eqnarray*}
with $\Delta'=x_3x_6-x_5x_4$, $x_1'=x_3+x_1x_4$, and
$x_2'=x_3+x_2x_4$.

Moreover, there are the following relations between the form
factors ($i=1...6$):
\begin{eqnarray*}
&&f_{2i}^\alpha=f^\alpha_{2i-1}(l^{\tilde{t}}_{ik}\leftrightarrow
k^{\tilde{t}}_{ik}, \ R^{\tilde{b}}_{j1}\leftrightarrow
R^{\tilde{b}}_{j2}, \ R^{\tilde{t}}_{i1}\leftrightarrow
R^{\tilde{t}}_{i2}),
\\
&&f_{2i}^{{\rm Box}(\beta)}= f^{{\rm
Box}(\beta)}_{2i-1}(l^{\tilde{t}}_{ik}\leftrightarrow
k^{\tilde{t}}_{ik}, \ R^{\tilde{b}}_{j1}\leftrightarrow
R^{\tilde{b}}_{j2}, \ R^{\tilde{t}}_{i1}\leftrightarrow
R^{\tilde{t}}_{i2}).
\end{eqnarray*}
Thus we will only present the explicit expressions of
$f^\alpha_{2i-1}$ and $f^{{\rm Box}(\beta)}_{2i-1}$.

For the diagram $(a)$ in Fig.2, we find
\begin{eqnarray*}
&&f_1^a=-3m_{\tilde{g}}C_0^fR^{\tilde{b}}_{j1} R^{\tilde{b}}_{j2}
k^{\tilde{t}}_{ik},
\\
&&f^a_3=\frac{1}{6s}l^{\tilde{t}}_{ik}\{(9-2\epsilon)B_0(s,0,0) -
40(1- 2 \epsilon)C_{00}^a + 18sC_0^a
+(R^{\tilde{b}}_{j1})^2[36C_{00}^f + 4C_{00}^g
\\ && \hspace{0.8cm}
-18B_0(s,m_{\tilde{g}}^2,m_{\tilde{b}_j}^2) ]\},
\\
&& f^a_5 =2f^a_3 + \frac{1}{3s}l^{\tilde{t}}_{ik}\{s(7C_1^a
-2C_0^a - 20(1-\epsilon)(C_{12}^a + C_{22}^a + C_2^a) +16C_2^a) +
(R^{\tilde{b}}_{j1})^2[2C_0^g(m_{\tilde{b}_j}^2
\\ && \hspace{0.8cm}
-m_{\tilde{g}}^2) + 2B_0(s,m_{\tilde{g}}^2,m_{\tilde{b}_j}^2) -
2B_0(0,m_{\tilde{b}_j}^2,m_{\tilde{b}_j}^2) + 2s(C_0^g + C_{11}^g
+ C_{12}^g + 2C_1^g + 9C_2^f + C_2^g
\\ && \hspace{0.8cm}
+ 9C_{12}^f + 9C_{22}^f)]\},
\\
&&f_{9}^a=-\frac{2}{s}f_1^a +\frac{2}{3s}m_{\tilde{g}}(C_1^g +
9C_2^f)R^{\tilde{b}}_{j1} R^{\tilde{b}}_{j2} k^{\tilde{t}}_{ik}.
\end{eqnarray*}
For the diagram $(b)$ in Fig.2, we find
\begin{eqnarray*}
&&f_1^b=\frac{4}{3}\{2l^{\tilde{b}}_{jk}R^{\tilde{b}}_{j1}[
m_tR^{\tilde{t}}_{i2} (C_1^j + C_2^j)
-m_{\tilde{g}}R^{\tilde{t}}_{i1} (C_0^j + C_1^j + C_2^j) ]
-m_{\tilde{\chi}_k^-} [l^{\tilde{t}}_{ik}(2C_0^b + C_1^b + 2C_2^b)
\\ && \hspace{0.8cm}
- 2k^{\tilde{b}}_{jk}R^{\tilde{b}}_{j1}
R^{\tilde{t}}_{i2}C_2^j]\},
\\
&&f_3^b=\frac{4}{3s}\{l^{\tilde{t}}_{ik}
[B_0(m_{\tilde{\chi}_k^-}^2,0,m_{\tilde{t}_i}^2) -2B_0(s,0,0) -
s_{\Delta}(2C_0^b + C_1^b + C_2^b) +m_{\tilde{t}_i}^2(C_1^b
-2C_0^b + C_2^b)
\\ && \hspace{0.8cm}
- m_{\tilde{\chi}_k^-}^2 C_2^b] +
2k^{\tilde{t}}_{jk}R^{\tilde{b}}_{j1} [m_tm_{\tilde{g}}
R^{\tilde{t}}_{i1} C_0^j +
R^{\tilde{t}}_{i2}(B_0(m_{\tilde{\chi}_k^-}^2,m_t^2,m_{\tilde{b}_j}^2)
+ m_{\tilde{g}}^2 C_0^j + m_{\tilde{t}_i}^2(C_1^j + C_2^j))]
\\ && \hspace{0.8cm}
-2(t_2+ u_2 + m_{\tilde{\chi}_k^-}^2)[l^{\tilde{t}}_{ik}(2C_0^b +
C_1^b + 2C_2^b) - 2k^{\tilde{b}}_{jk}R^{\tilde{b}}_{j1}
R^{\tilde{t}}_{i2}C_2^j]
+2m_{\tilde{\chi}_k^-}l^{\tilde{b}}_{jk}R^{\tilde{b}}_{j1}
[m_{\tilde{g}}R^{\tilde{t}}_{i1} (C_0^j
\\ && \hspace{0.8cm}
+ C_1^j) - m_tR^{\tilde{t}}_{i2} C_1^j ]\},
\\
&&f_5^b=2f_3^b.
\end{eqnarray*}
For the diagram $(c)$ in Fig.2, we find
\begin{eqnarray*}
&& f_1^c = -\frac{8}{3s} l^{\tilde{t}}_{ik} m_{\tilde{g}}
R_{j1}^{\tilde{b}} R_{j2}^{\tilde{b}} B_0(s,m_{\tilde{g}}^2,
m_{\tilde{b}_j}^2),
\\
&& f_3^c=-\frac{4}{3s} l^{\tilde{t}}_{ik}[2(1- \epsilon)B_1(s,0,0)
+2 (R_{j1}^{\tilde{b}})^2B_1(s,m_{\tilde{g}}^2,
m_{\tilde{b}_j}^2)],
\\
&& f_5^c= 2f_3^c.
\end{eqnarray*}
For the diagram $(d)$ in Fig.2, we find
\begin{eqnarray*}
&& f_5^d=\frac{1}{12}
\{\frac{l_{ik}^{\tilde{t}}}{t_1}[16B_0(t,0,m_{\tilde{t}_i}^2) +
16B_0(m_{\tilde{t}_i}^2,0,m_{\tilde{t}_i}^2) + t_1(18C_0^d -
2C_1^c + 27C_2^d) - 8m_{\tilde{t}_i}^2(C_1^c - 9C_2^d)]
\\ && \hspace{0.8cm}
- \frac{4l_{jk}^{\tilde{t}}R^{\tilde{t}}_{i1}}{t -m_{\tilde{t}_j}
}[8R^{\tilde{t}}_{j1}(B_0(t,m_{\tilde{g}}^2,m_t^2) +
B_0(m_{\tilde{t}_i}^2,m_{\tilde{g}}^2,m_t^2)) -
(R^{\tilde{t}}_{j1}(2m_{\tilde{g}}^2 + 2m_t^2 - t
-m_{\tilde{t}_i}^2)
\\ && \hspace{0.8cm}
- 4m_{\tilde{g}}m_tR^{\tilde{t}}_{j2})(9C_1^h - C_1(t,
m_{\tilde{t}_i}^2, 0; m_t^2, m_{\tilde{g}}^2, m_t^2))]\}
+(R^{\tilde{t}}_{i1}\leftrightarrow R^{\tilde{t}}_{i2},\
R^{\tilde{t}}_{j1}\leftrightarrow R^{\tilde{t}}_{j2}),
\\
&& f_7^d=-f_5^d.
\end{eqnarray*}
For the diagram $(e)$ in Fig.2, we find
\begin{eqnarray*}
&&f_5^e=\frac{8}{3t_1}\{l_{ik}^{\tilde{t}}
[B_0(m_{\tilde{\chi}_k^-}^2,0,m_{\tilde{t}_i}^2) - t_2(C_0^e +
C_1^e + C_2^e) - 2m_{\tilde{t}_i}^2C_0^e - 2m_{\tilde{\chi}_k^-}^2
(C_1^e + C_2^e)]
\\ && \hspace{1.2cm}
- 2R^{\tilde{t}}_{i1} R^{\tilde{b}}_{j1}[k_{jk}^{\tilde{b}}
(B_0(m_{\tilde{\chi}_k^-}^2,m_t^2,m_{\tilde{b}_j}^2) +
m_{\tilde{g}}^2C_0^i + t_2C_1^i + m_{\tilde{\chi}_k^-}^2 C_1^i) -
l_{jk}^{\tilde{b}}m_{\tilde{\chi}_k^-} m_{\tilde{g}}(C_0^i +
C_1^i)]
\\ && \hspace{1.2cm}
+ 2R^{\tilde{t}}_{i2}R^{\tilde{b}}_{j1}m_t (k_{jk}^{\tilde{b}}
m_{\tilde{g}} C_0^i -l_{jk}^{\tilde{b}}m_{\tilde{\chi}_k^-}
C_1^i)\},
\\
&& f_7^e=-f_5^e.
\end{eqnarray*}
For the diagram $(f)$ in Fig.2, we find
\begin{eqnarray*}
&& f_5^f= -\frac{8}{3t_1} \{\frac{l_{ik}^{\tilde{t}}}{t_1}[2(t +
m_{\tilde{t}_i}^2 ) B_0(t,0,m_{\tilde{t}_i}^2)
-A_0(m_{\tilde{t}_i}^2)] + \frac{4l_{jk}^{\tilde{t}}} {t
-m_{\tilde{t}_j}^2} [(tB_1 +A_0(m_t^2) +m_{\tilde{g}}^2 B_0)
(R_{i1}^{\tilde{t}} R_{j1}^{\tilde{t}}
\\ && \hspace{0.8cm}
+ R_{i2}^{\tilde{t}}R_{j2}^{\tilde{t}}) -m_t m_{\tilde{g}}
(R_{i2}^{\tilde{t}} R_{j1}^{\tilde{t}} + R_{i1}^{\tilde{t}}
R_{j2}^{\tilde{t}})B_0](t,m_{\tilde{g}}^2,m_t^2)\},
\\
&& f_7^f= -f_5^f.
\end{eqnarray*}
For the box diagram $(a)$ in Fig.3, we find
\begin{eqnarray*}
&& f_1^{{\rm Box}(a)}=
\frac{3}{2t_2}l_{ik}^{\tilde{t}}m_{\tilde{\chi}_k^-}
[B_0(m_{\tilde{\chi}_k^-}^2,0,m_{\tilde{t}_i}^2) -
B_0(t,0,m_{\tilde{t}_i}^2) + t_2(C_1^b + 4D_{00}^a - 2t_1D_1^a)],
\\
&& f_3^{{\rm Box}(a)}= -\frac{3}{2}l_{ik}^{\tilde{t}} [ C_0^a +
C_0^b + C_1^a + C_1^b + C_2^a + C_2^b + 4D_{00}^a - 2t_1(D_0^a +
D_1^a + D_2^a + D_3^a)],
\\
&& f_5^{{\rm Box}(a)}= \frac{3}{2t_2}l_{ik}^{\tilde{t}}
\{2B_0(m_{\tilde{\chi}_k^-}^2,0,m_{\tilde{t}_i}^2) -
2B_0(t,0,m_{\tilde{t}_i}^2) - t_2[2C_0^a + 2C_0^e + C_1^b + 2C_1^e
- 2C_2^a + C_2^b
\\ && \hspace{1.2cm}
+ 6D_{00}^a + s(D_{11}^a  +5D_{13}^a + 2D_1^a + 4D_{33}^a +4D_3^a)
- s_{\Delta}(D_{11}^a + 2D_{13}^a + 2D_1^a + D_{33}^a + 2D_3^a)
\\ && \hspace{1.2cm}
- t_1(2D_0^a + 3D_{12}^a + 3D_1^a + 3D_{23}^a + 2D_2^a + 3D_3^a) +
u_2(2D_0^a + D_{11}^a - 3D_{12}^a + D_{13}^a + 3D_1^a
\\ && \hspace{1.2cm}
- 3D_{23}^a - 2D_2^a + D_3^a)] - t_2m_{\tilde{t}_i}^2 (2D_{11}^a +
D_{13}^a - 4D_1^a - D_{33}^a - 6D_3^a) +
t_2m_{\tilde{\chi}_k^-}^2(D_{13}^a + D_{33}^a
\\ && \hspace{1.2cm}
- 2D_3^a)\},
\\
&& f_7^{{\rm Box}(a)}= -\frac{3}{2t_2}l_{ik}^{\tilde{t}}
\{4B_0(m_{\tilde{\chi}_k^-}^2,0,m_{\tilde{t}_i}^2) -
4B_0(t,0,m_{\tilde{t}_i}^2) + t_2[2C_0^b - 4C_0^e + C_1^b -
s(4D_{13}^a + 2D_1^a
\\ && \hspace{1.2cm}
+D_{11}^a) + s_{\Delta}(D_{13}^a + 2D_1^a - 2D_3^a +D_{11}^a) -
t_1(D_1^a - 2D_2^a) - u_2(3D_1^a - 2D_2^a +D_{11}^a)
\\ && \hspace{1.2cm}
- (2D_0^a - 3D_{12}^a)(t_1 + u_2)] -
t_2m_{\tilde{t}_i}^2(2D_{11}^a - D_{13}^a - 4D_1^a + 2D_3^a) +
t_2m_{\tilde{\chi}_k^-}^2(D_{13}^a - 2D_3^a)\},
\\
&& f_9^{{\rm Box}(a)}= 6 l_{ik}^{\tilde{t}}m_{\tilde{\chi}_k^-}
(D_{12}^a + D_{13}^a + D_{23}^a + D_{33}^a + D_3^a),
\\
&& f_{11}^{{\rm Box}(a)}= - 6
l_{ik}^{\tilde{t}}m_{\tilde{\chi}_k^-} (D_{12}^a + D_{13}^a).
\end{eqnarray*}
For the box diagram $(b)$ in Fig.3, we find
\begin{eqnarray*}
&& f_{1}^{{\rm Box}(b)} =
-\frac{1}{6}l_{ik}^{\tilde{t}}m_{\tilde{\chi}_k^-} (2C_0^b + C_1^b
- 4D_{00}^b),
\\
&& f_{3}^{{\rm Box}(b)} = \frac{1}{6}l_{ik}^{\tilde{t}}[2C_0^a +
2C_0^b - C_0(u,m_{\tilde{\chi}_k^-}^2, 0; 0, m_{\tilde{t}_i}^2, 0)
+ C_1^b + C_2^b - 4D_{00}^b + (t +m_{\tilde{\chi}_k^-}^2) D_3^b
\\ && \hspace{1.2cm}
- 2u_1 D_0^b -(u_1 +2m_{\tilde{t}_i}^2) (D_1^b + D_3^b)],
\\
&& f_{5}^{{\rm Box}(b)} = \frac{1}{3}l_{ik}^{\tilde{t}} [2(C_0^a +
C_1^a + C_2^a - D_{00}^b) + (s + t_2)(2D_0^b + D_{11}^b + D_{12}^b
+ 3D_1^b + 2D_2^b + D_3^b)
\\ && \hspace{1.2cm}
- s( D_{23}^b  + D_{33}^b ) +  t_2( D_{13}^b + D_3^b) -
2m_{\tilde{t}_i}^2(D_{11}^b + D_{12}^b + D_{13}^b + D_1^b)],
\\
&& f_{7}^{{\rm Box}(b)} = \frac{1}{3}l_{ik}^{\tilde{t}}[2C_0^b +
C_1^b + C_1(u,m_{\tilde{\chi}_k^-}^2, 0; 0, m_{\tilde{t}_i}^2, 0)
+ u_1(D_{11}^b + 2D_1^b) + 2m_{\tilde{t}_i}^2D_{11}^b],
\\
&& f_{9}^{{\rm Box}(b)} = \frac{2}{3}l_{ik}^{\tilde{t}}
m_{\tilde{\chi}_k^-}(D_{13}^b + D_{23}^b + D_{33}^b + D_3^b),
\\
&& f_{11}^{{\rm Box}(b)} = -\frac{2}{3}l_{ik}^{\tilde{t}}
m_{\tilde{\chi}_k^-}D_{13} ^b.
\end{eqnarray*}
For the box diagram $(c)$ in Fig.3, we find
\begin{eqnarray*}
&& f_1^{{\rm Box}(c)} =- m_{\tilde{\chi}_k^-} f_3^{{\rm Box}(c)} =
- \frac{2}{3}m_{\tilde{\chi}_k^-} l_{ik}^{\tilde{t}} D_{00}^c,
\\
&& f_5^{{\rm Box}(c)} = - \frac{1}{3t_2}l_{ik}^{\tilde{t}}
\{2B_0(m_{\tilde{\chi}_k^-}^2,0,m_{\tilde{t}_i}^2) -
2B_0(t,0,m_{\tilde{t}_i}^2) + t_2[-2C_0^e - 2C_1^e + 2D_{00}^c +
(s +2t
\\ && \hspace{1.2cm}
+ 2 u_2)(D_{13}^c + D_1^c) + 2u (D_{13}^c + D_3^c + D_{23}^c +
D_{33}^c) + u_1(2D_0^c + 3D_1^c - D_{23}^c + 2D_2^c
\\ && \hspace{1.2cm}
- D_{33}^c + D_3^c) + D_{12}^c( t +m_{\tilde{\chi}_k^-}^2 - 2u_2)
+ D_{11}^c(t + u + u_2)]\},
\\
&& f_7^{{\rm Box}(c)} =  \frac{1}{3t_2}l_{ik}^{\tilde{t}}
\{2B_0(m_{\tilde{\chi}_k^-}^2,0,m_{\tilde{t}_i}^2) -
2B_0(t,0,m_{\tilde{t}_i}^2) + t_2[-2C_0^e + sD_1^c + (t_1 +u_2)
(2D_1^c +D_{11}^c
\\ && \hspace{1.2cm}
+D_{13}^c) + (C_0 + C_1 + C_2)(m_{\tilde{\chi}_k^-}^2,0,u;
0,m_{\tilde{t}_i}^2,m_{\tilde{t}_i}^2) + u_1(2D_0^c + 3D_1^c +
D_{11}^c+ 2D_{13}^c
\\ && \hspace{1.2cm}
+ D_{33}^c + 3D_3^c) + 2m_{\tilde{\chi}_k^-}^2 (D_{11}^c +
2D_{13}^c + D_1^c + D_{33}^c + D_3^c)]\},
\\
&& f_9^{{\rm Box}(c)} =  - \frac{2}{3}l_{ik}^{\tilde{t}}
m_{\tilde{\chi}_k^-}(D_{11}^c + D_{12}^c + D_{13}^c + D_1^c),
\\
&& f_{11}^{{\rm Box}(c)} =  \frac{2}{3}l_{ik}^{\tilde{t}}
m_{\tilde{\chi}_k^-} (D_{11}^c + D_{13}^c + D_1^c).
\end{eqnarray*}
For the box diagram $(d)$ in Fig.3, we find
\begin{eqnarray*}
&& f_{1}^{{\rm Box}(d)} = 3R^{\tilde{b}}_{j1}
\{k^{\tilde{b}}_{jk}m_{\tilde{\chi}_k^-} R^{\tilde{t}}_{i2} (C_0^j
+ C_1^i - 2D_{00}^d + t_1D_2^d) + l^{\tilde{b}}_{jk}[m_{\tilde{g}}
R^{\tilde{t}}_{i1} (C_0^i - C_0^j + 2D_{00}^d + t_1D_0^d
\\ && \hspace{1.2cm}
+ u_2D_0^d + u_2D_2^d) + m_t R^{\tilde{t}}_{i2} (C_0^j - 2D_{00}^d
- u_2D_2^d)]\},
\\
&& f_{3}^{{\rm Box}(d)} =  3R^{\tilde{b}}_{j1} \{
l^{\tilde{b}}_{jk}m_{\tilde{\chi}_k^-}[m_{\tilde{g}}R^{\tilde{t}}_{i1}
(D_0^d + D_2^d) - m_tR_{i2}^{\tilde{t}}D_2^d] +
k^{\tilde{b}}_{jk}[m_t m_{\tilde{g}}R_{i1}^{\tilde{t}} D_0^d -
R_{i2}^{\tilde{t}}(C_0^j - 2D_{00}^d
\\ && \hspace{1.2cm}
+ m_{\tilde{g}}^2D_0^d + tD_2^d)]\},
\\
&& f_{5}^{{\rm Box}(d)} =  -6 R^{\tilde{b}}_{j1}
\{l^{\tilde{b}}_{jk}m_{\tilde{\chi}_k^-}[m_{\tilde{g}}R^{\tilde{t}}_{i1}
(D_{12}^d + D_1^d + D_{22}^d + D_2^d) - m_tR_{i2}^{\tilde{t}}
(D_{12}^d + D_{22}^d)]
\\ && \hspace{1.2cm}
+ k^{\tilde{b}}_{jk}[m_t m_{\tilde{g}}R_{i1}^{\tilde{t}} (D_1^d +
D_2^d) + R_{i2}^{\tilde{t}}(C_0^j - 2D_{00}^d -
m_{\tilde{g}}^2(D_1^d + D_2^d) + (t_1 +u_2)(D_{11}^d + D_{12}^d
\\ && \hspace{1.2cm}
+ D_{13}^d + D_1^d + D_{23}^d) + t_1D_2^d -
m_{\tilde{t}_i}^2(D_{12}^d + D_{22}^d))]\},
\\
&& f_{7}^{{\rm Box}(d)} =  6 R^{\tilde{b}}_{j1}
\{l^{\tilde{b}}_{jk}m_{\tilde{\chi}_k^-}
[m_{\tilde{g}}R^{\tilde{t}}_{i1} (D_{22}^d + D_2^d) -
m_tR_{i2}^{\tilde{t}} D_{22}^d] + k^{\tilde{b}}_{jk}[m_t
m_{\tilde{g}}R_{i1}^{\tilde{t}} D_2^d - R_{i2}^{\tilde{t}}(C_1^i
\\ && \hspace{1.2cm}
+ m_{\tilde{g}}^2 D_2^d - t_1(D_{12}^d + D_{23}^d) - u_2(D_{12}^d
+ D_{23}^d) + m_{\tilde{t}_i}^2 D_{22}^d)]\},
\\
&& f_{9}^{{\rm Box}(d)} =  6 R^{\tilde{b}}_{j1}
\{l^{\tilde{b}}_{jk}[m_{\tilde{g}}R^{\tilde{t}}_{i1}(D_0^d +
D_{11}^d + 2D_{12}^d + D_{13}^d + 2D_1^d + D_{22}^d + D_{23}^d +
2D_2^d)
\\ && \hspace{1.2cm}
- m_tR_{i2}^{\tilde{t}}(D_{11}^d + 2D_{12}^d + D_{13}^d + D_1^d +
D_{22}^d + D_{23}^d + D_2^d)] -
k^{\tilde{b}}_{jk}m_{\tilde{\chi}_k^-} R_{i2}^{\tilde{t}}
(D_{11}^d
\\ && \hspace{1.2cm}
+ D_{12}^d + D_{13}^d + D_1^d + D_{23}^d) \},
\\
&& f_{11}^{{\rm Box}(d)} =  6 R^{\tilde{b}}_{j1}\{
l^{\tilde{b}}_{jk}[m_tR_{i2}^{\tilde{t}} (D_{12}^d + D_{22}^d +
D_{23}^d) - m_{\tilde{g}}R^{\tilde{t}}_{i1}(D_{12}^d + D_{22}^d +
D_{23}^d + D_2^d)]
\\ && \hspace{1.2cm}
+ k^{\tilde{b}}_{jk}m_{\tilde{\chi}_k^-} R_{i2}^{\tilde{t}}
(D_{12}^d + D_{23}^d)\}.
\end{eqnarray*}
For the box diagram $(e)$ in Fig.3, we find
\begin{eqnarray*}
&& f_{1}^{{\rm Box}(e)} = - \frac{2}{3} R_{j1}^{\tilde{b}}
[k^{\tilde{b}}_{jk}m_{\tilde{\chi}_k^-} R_{i2}^{\tilde{t}} +
l_{jk}^{\tilde{b}} (m_t R^{\tilde{t}}_{i2}
-m_{\tilde{g}}R_{i1}^{\tilde{t}})] D_{00}^e,
\\
&& f_{3}^{{\rm Box}(e)} =  \frac{2}{3} k^{\tilde{b}}_{jk}
R_{j1}^{\tilde{b}} R_{i2}^{\tilde{t}} D_{00}^e,
\\
&& f_{5}^{{\rm Box}(e)} =  \frac{2}{3} R_{j1}^{\tilde{b}}
\{l^{\tilde{b}}_{jk} m_{\tilde{\chi}_k^-} [ (R^{\tilde{t}}_{i2}
m_t - m_{\tilde{g}} R_{i1}^{\tilde{t}})(D_{22}^e + D_{12}^e +
D_{23}^e + D_2^e ) - m_{\tilde{g}} R_{i1}^{\tilde{t}} (D_0^e +
D_1^e
\\ && \hspace{1.2cm}
+ D_2^e + D_3^e)] + k^{\tilde{b}}_{jk} [R_{i2}^{\tilde{t}}
(2D_{00}^e + m_{\tilde{t}_i}^2 (D_{12}^e + D_{22}^e + D_{23}^e +
D_2^e)) + s( D_{12}^e + D_{11}^e
\\ && \hspace{1.2cm}
+ D_{13}^e + D_1^e ) + m_{\tilde{g}}(R_{i2}^{\tilde{t}}
m_{\tilde{g}} - R^{\tilde{t}}_{i1} m_t)(D_0^e + D_1^e + D_2^e +
D_3^e) ] \},
\\
&& f_{7}^{{\rm Box}(e)} =  \frac{2}{3} R_{j1}^{\tilde{b}}
\{l^{\tilde{b}}_{jk} m_{\tilde{\chi}_k^-} [m_{\tilde{g}}
R_{i1}^{\tilde{t}} (D_{22}^e + D_2^e) - m_t R^{\tilde{t}}_{i2}
D_{22}^e] +k^{\tilde{b}}_{jk} [m_{\tilde{g}}m_t R^{\tilde{t}}_{i1}
D_2^e + R_{i2}^{\tilde{t}} ((C_0
\\ && \hspace{1.2cm}
+ C_1 + C_2)(m_{\tilde{\chi}_k^-}^2,0,u; m_t^2,
m_{\tilde{b}_j}^2,m_{\tilde{b}_j}^2) - m_{\tilde{g}}^2 D_2^e -
sD_{12}^e - m_{\tilde{t}_i}^2D_{22}^e)] \},
\\
&& f_{9}^{{\rm Box}(e)} =  \frac{2}{3} R_{j1}^{\tilde{b}}
\{l^{\tilde{b}}_{jk}[(m_{\tilde{g}} R_{i1}^{\tilde{t}} - m_t
R^{\tilde{t}}_{i2})(D_{11}^e + 2D_{12}^e + D_{13}^e + D_1^e +
D_{22}^e + D_{23}^e + D_2^e)
\\ && \hspace{1.2cm}
+ m_{\tilde{g}} R_{i1}^{\tilde{t}} (D_0^e + D_1^e + D_2^e +
D_3^e)] - k^{\tilde{b}}_{jk} m_{\tilde{\chi}_k^-}
R^{\tilde{t}}_{i2} (D_{11}^e + D_{12}^e + D_{13}^e + D_1^e)\},
\\
&& f_{11}^{{\rm Box}(e)} =  \frac{2}{3} R_{j1}^{\tilde{b}}
\{l^{\tilde{b}}_{jk}[m_t R^{\tilde{t}}_{i2}(D_{12}^e + D_{22}^e) -
m_{\tilde{g}} R_{i1}^{\tilde{t}}(D_{12}^e + D_{22}^e + D_2^e)]
+k^{\tilde{b}}_{jk} m_{\tilde{\chi}_k^-} R^{\tilde{t}}_{i2}
D_{12}^e \}.
\end{eqnarray*}
For the box diagram $(f)$ in Fig.3, we find
\begin{eqnarray*}
&& f_{1}^{{\rm Box}(f)} = \frac{1}{3} R_{j1}^{\tilde{b}}\{
l^{\tilde{b}}_{jk} [(m_t R^{\tilde{t}}_{i2} - m_{\tilde{g}}
R^{\tilde{t}}_{i1})(C_0^k - 2D_{00}^f + s(D_0^f + D_1^f + D_2^f +
D_3^f)) - m_t R^{\tilde{t}}_{i2}(C_0^i
\\ && \hspace{1.2cm}
+ s D_0^f + t_1D_0^f)] + k^{\tilde{b}}_{jk} m_{\tilde{\chi}_k^-}
R^{\tilde{t}}_{i2}[C_0^k + C_1(t,0,m_{\tilde{\chi}_k^-}^2; m_t^2,
m_{\tilde{g}}^2, m_{\tilde{b}_j}^2) + C_2^i - 2D_{00}^f + sD_2^f
\\ && \hspace{1.2cm}
- t_1(D_0^f + D_1^f + D_3^f)]\},
\\
&& f_{3}^{{\rm Box}(f)} = \frac{1}{3} R_{j1}^{\tilde{b}}\{
l^{\tilde{b}}_{jk} m_{\tilde{\chi}_k^-}[m_{\tilde{g}}
R^{\tilde{t}}_{i1}(D_0^f + D_1^f + D_3^f) - m_t R^{\tilde{t}}_{i2}
(D_1^f + D_3^f)] + k^{\tilde{b}}_{jk} [R^{\tilde{t}}_{i2} (C_0^i -
C_0^k
\\ && \hspace{1.2cm}
- C_0(0,m_{\tilde{\chi}_k^-}^2,u; m_t^2, m_t^2, m_{\tilde{b}_j}^2)
+ 2D_{00}^f - m_{\tilde{g}}^2D_0^f - sD_2^f + t_1(D_0^f + D_3^f) -
m_{\tilde{t}_i}^2(D_1^f + D_3^f))
\\ && \hspace{1.2cm}
+ m_{\tilde{g}} m_t R^{\tilde{t}}_{i1} D_0^f]\},
\\
&& f_{5}^{{\rm Box}(f)} = \frac{2}{3} R_{j1}^{\tilde{b}}\{
l^{\tilde{b}}_{jk} m_{\tilde{\chi}_k^-}[(m_t R^{\tilde{t}}_{i2} -
m_{\tilde{g}} R^{\tilde{t}}_{i1})(D_{11}^f + D_{12}^f + 2D_{13}^f
+ D_1^f + D_{23}^f + D_{33}^f + D_3^f)
\\ && \hspace{1.2cm}
- m_{\tilde{g}} R^{\tilde{t}}_{i1} (D_0^f + D_1^f + D_2^f +
D_3^f)] - k^{\tilde{b}}_{jk} [m_{\tilde{g}} (m_t
R^{\tilde{t}}_{i1} - m_{\tilde{g}}R^{\tilde{t}}_{i2})(D_0^f +
D_1^f + D_2^f + D_3^f)
\\ && \hspace{1.2cm}
+ R^{\tilde{t}}_{i2}(C_0^i + C_1^i +
C_1(t,0,m_{\tilde{\chi}_k^-}^2; m_t^2, m_{\tilde{g}}^2,
m_{\tilde{b}_j}^2) + C_2^i - 2D_{00}^f + (s -
m_{\tilde{t}_i}^2)(D_{11}^f + D_{12}^f
\\ && \hspace{1.2cm}
+ D_{13}^f + D_1^f) - m_{\tilde{t}_i}^2 (D_{13}^f + D_{23}^f +
D_{33}^f + D_3^f))]\},
\\
&& f_{7}^{{\rm Box}(f)} = \frac{2}{3} R_{j1}^{\tilde{b}}\{
l^{\tilde{b}}_{jk} m_{\tilde{\chi}_k^-}[m_{\tilde{g}}
R^{\tilde{t}}_{i1} (D_0^f + D_1^f + D_3^f) + (m_{\tilde{g}}
R^{\tilde{t}}_{i1}- m_t R^{\tilde{t}}_{i2} )(D_{11}^f + 2D_{13}^f
\\ && \hspace{1.2cm}
+ D_1^f + D_{33}^f + D_3^f)] + k^{\tilde{b}}_{jk} [(m_{\tilde{g}}
m_t R^{\tilde{t}}_{i1} - R^{\tilde{t}}_{i2}(m_{\tilde{g}}^2 -
t_1))(D_0^f + D_1^f + D_3^f)
\\ && \hspace{1.2cm}
- R^{\tilde{t}}_{i2}(\frac{1}{u_2}
(B_0(m_{\tilde{\chi}_k^-}^2,m_t^2,m_{\tilde{b}_j}^2) - B_0(u,
m_t^2,m_{\tilde{b}_j}^2)) - C_0^i + C_0^k - C_1^i - s(D_{11}^f +
D_{13}^f
\\ && \hspace{1.2cm}
+ D_1^f - D_2^f) + m_{\tilde{t}_i}^2 (D_{11}^f + 2D_{13}^f + D_1^f
+ D_{33}^f + D_3^f))] \},
\\
&& f_{9}^{{\rm Box}(f)} = \frac{2}{3} R_{j1}^{\tilde{b}}\{
l^{\tilde{b}}_{jk} [m_{\tilde{g}} R^{\tilde{t}}_{i1}(D_0^f + D_1^f
+ D_2^f + D_3^f) + (m_{\tilde{g}} R^{\tilde{t}}_{i1} - m_t
R^{\tilde{t}}_{i2}) (D_{13}^f + D_{23}^f + D_{33}^f
\\ && \hspace{1.2cm}
+ D_3^f)] +k^{\tilde{b}}_{jk} m_{\tilde{\chi}_k^-}
R^{\tilde{t}}_{i2}(D_{11}^f + D_{12}^f + D_{13}^f + D_1^f) \},
\\
&& f_{11}^{{\rm Box}(f)} = \frac{2}{3} R_{j1}^{\tilde{b}}\{
l^{\tilde{b}}_{jk} [m_t R^{\tilde{t}}_{i2}(D_{13}^f - D_1^f +
D_{33}^f) - m_{\tilde{g}} R^{\tilde{t}}_{i1}(D_{13}^f + D_{33}^f +
D_3^f)]
\\ && \hspace{1.2cm}
- k^{\tilde{b}}_{jk} m_{\tilde{\chi}_k^-}
R^{\tilde{t}}_{i2}(D_{11}^f + D_{13}^f + D_1^f)\}.
\end{eqnarray*}
For the box diagram $(g)$ in Fig.3, we find
\begin{eqnarray*}
&& f_{1}^{{\rm Box}(g)} = \frac{7}{6t_2}l_{ik}^{\tilde{t}}
m_{\tilde{\chi}_k^-}
(B_0(m_{\tilde{\chi}_k^-}^2,0,m_{\tilde{t}_i}^2) -
B_0(t,0,m_{\tilde{t}_i}^2)),
\\
&& f_5^{{\rm Box}(g)} =  - \frac{7}{3}l_{ik}^{\tilde{t}} C_2^e.
\end{eqnarray*}
%%%%%%%%%%%%%%%%%%%%%%%%%%%% appendix B %%%%%%%%%%%%%%%%%%%%%%%
\section{}
In this Appendix, we collect the explicit expressions for the
amplitudes squared of the radiations of a real gluon and a
massless (anti-)quark. Since these expressions are only used to
the hard non-collinear parts of real corrections in
Eq.(\ref{nonHC}) and Eq.(\ref{qHC}), they have no singularities
and can be calculated in $n=4$ dimensions.

For the real gluon emission, we find
\begin{eqnarray*}
&&\overline{|M^{gb}_{ik}|}^2=X_{ik}[\overline{|M^{(s)}|}^2+
\overline{|M^{(t)}|}^2 +\overline{|M^{(st)}|}^2] + (s
\leftrightarrow t', \ s_4 \leftrightarrow u_7, \ s_{3\Delta}
\leftrightarrow u_2 ),
\end{eqnarray*}
where $X_{ik}=g^2g_s^4(|l^{\tilde{t}}_{ik}|^2 +
|k^{\tilde{t}}_{ik}|^2)$, and
\begin{eqnarray*}
&&\overline{|M^{(s)}|}^2=\frac{1}{ss_5}\{\frac{1}{36s_4t'}
[t_2u'm_{\tilde{t}_i}^2  -s_4u_6(t_2 +u_2)+ u_6(u_6(s_{3\Delta} -
u_2) + (s_5 +u_7)(s_{3\Delta} + t_2)
\\ && \hspace{1.5cm}
- u'(s_{5\Delta} + m_{\tilde{\chi}_k^-}^2)) -s_4
t'm_{\tilde{\chi}_k^-}^2] -\frac{1}{8s_4u'}[s_{5\Delta}t'u_6 +
u_6^2(u_2 -s_{3\Delta}) - u_6u_7(s_{3\Delta} + t_2
\\ && \hspace{1.5cm}
+ u_2) +(s_4t_2-ss_{5\Delta})(u_6 +u_7) + t_2u_7^2 -
m_{\tilde{t}_i}^2(u_2(2s+t') +4t_2t'-3ss_{3\Delta})
\\ && \hspace{1.5cm}
- m_{\tilde{\chi}_k^-}^2(2u_6(t'+s) + u_7(4s +t'))]
-\frac{t'}{8s_5u'}[\Delta_t(4t_2 + u_2) + (4u_6 + u_7)(
2m_{\tilde{\chi}_k^-}^2 + s_{5\Delta})]
\\ && \hspace{1.5cm}
-\frac{u'}{36s_5t'}[\Delta_tt_2 + u_6(s_{5\Delta} +
2m_{\tilde{\chi}_k^-}^2)] -\frac{1}{72s_4}[11s_{5\Delta}u_6
-m_{\tilde{t}_i}^2(31t_2 + 34u_2) - m_{\tilde{\chi}_k^-}^2(9u_6
\\ && \hspace{1.5cm}
+ 34u_7)] -\frac{1}{72s_5}[\Delta_t(31t_2 + 16u_2) + (31u_6 +
16u_7)(s_{5\Delta} + 2m_{\tilde{\chi}_k^-}^2)]\}
\\ && \hspace{1.5cm}
+\frac{1}{s_5^2}\{\frac{1}{8u'^2} [4s(\Delta_ts_{3\Delta} +
s_4s_{5\Delta} + 2s_4m_{\tilde{\chi}_k^-}^2) +u'\Delta_t(u_2
-2t_2) + u'(s_4 - 2u_6)(s_{5\Delta}
\\ && \hspace{1.5cm}
+ 2m_{\tilde{\chi}_k^-}^2)
+s_5u'(s_{5\Delta} -m_{\tilde{\chi}_k^-}^2)]\}
+\frac{4u_2}{9ss_4^2}m_{\tilde{t}_i}^2,
\\
&&\overline{|M^{(t)}|}^2=\frac{1}{t_1^2}\{\frac{1}{16u'^2}[2t_2
(s_{3\Delta} - s_4 + t')(s + u_2 - u_7) +4t_2u'(s_4 + s_{5\Delta}
+ u_6 - m_{\tilde{t}_i}^2) +s_{5\Delta}t_1u']
\\ && \hspace{1.5cm}
-\frac{t_2}{4s_4u'}[u_7(s_{3\Delta} + s_{5\Delta} + t' + u_6) -
m_{\tilde{t}_i}^2(s - s_{3\Delta} - t' + u_2 + u_7)]
+\frac{1}{36s_4u_7}[t_2(4m_{\tilde{\chi}_k^-}^4
\\ && \hspace{1.5cm}
+s_4u_7) +2s_{5\Delta}t_1m_{\tilde{t}_i}^2]
-\frac{1}{18s_4^2}[m_{\tilde{t}_i}^2( t_2(16m_{\tilde{t}_i}^2
+7s_4) +8 t_1(s_{3\Delta} +s_{5\Delta}))
-\frac{9}{4}s_4s_{5\Delta}t_1]\}
\\ && \hspace{1.5cm}
-\frac{1}{16s_4t_1u'} [t_2u_7 + 4s_{3\Delta}m_{\tilde{t}_i}^2 +
4u_7m_{\tilde{\chi}_k^-}^2 - s_{5\Delta}(s + s_{3\Delta} - t' +
u_2 + u_7) - u_6(2s_{3\Delta}
\\ && \hspace{1.5cm}
- 2t_2 - u_2)],
\\
&&\overline{|M^{(st)}|}^2=
\frac{1}{36s_5t_1}\{\frac{9}{s_4u'}[u_7(s_{5\Delta}t' -
s_{3\Delta}u_6 + 2m_{\tilde{\chi}_k^-}^2 (t' + u_6)) +
m_{\tilde{t}_i}^2(u_6(u_2 -s_{3\Delta}) + t_2u_7
\\ && \hspace{1.5cm}
+ (s_{5\Delta}+2m_{\tilde{\chi}_k^-}^2 )(t' -s ))] +
\frac{1}{2ss_4}[16m_{\tilde{t}_i}^2(u_2(t' + u_6) + t_2(2t' + 2u_6
+ u_7 + u'))
\\ && \hspace{1.5cm}
+ 16m_{\tilde{\chi}_k^-}^2 (t'(2u_6 + u_7) + u_6(2u_6 + 2u_7 +
u')) -s_4t_2(9s_{5\Delta} + 2u_6 - 7u_7 - 7m_{\tilde{t}_i}^2)
\\ && \hspace{1.5cm}
+ 7s_4u_2u_6 +ss_4(7s_{5\Delta} -2m_{\tilde{\chi}_k^-}^2)
-(9s_5(s_{5\Delta}t_2 + 3t_2u_6 + u_2u_6 + t_2u_7) +
14s_5t_2m_{\tilde{t}_i}^2
\\ && \hspace{1.5cm}
+ s_4m_{\tilde{\chi}_k^-}^2(9t_2 - 14u_6)-2su_6(s_{5\Delta} +
2m_{\tilde{\chi}_k^-}^2) - 2sm_{\tilde{t}_i}^2(8s_{3\Delta} +
7s_{5\Delta} + 14m_{\tilde{\chi}_k^-}^2)]
\\ && \hspace{1.5cm}
-\frac{9}{4su'}[2t'(\Delta_t +s_{5\Delta})(2t_2 + u_2) +
(s_{3\Delta} - s_4 - 2t_2 - t' - u_2 + 2u_6 + u_7)(t_2u_7
\\ && \hspace{1.5cm}
-u_2u_6) + (s_5t_2- 4t'm_{\tilde{\chi}_k^-}^2)(2u_6 + u_7)
-3ss_4t_2 - 2s\Delta_t(2t_2 + u_2) - 2ss_{5\Delta}(t_2
\\ && \hspace{1.5cm}
+ u_2 - u_6) +4sm_{\tilde{\chi}_k^-}^2(s_4 + 3u_6) + (s_5 -s)
u_2u_6] -\frac{9}{u'^2}[ss_{5\Delta}(s_{3\Delta} - s_4 + t') + (s
\\ && \hspace{1.5cm}
+ u_2)(s_4t_2 + s_{3\Delta}u_6) + s_4(u_2u_6 - t_2u_7) +
2t'm_{\tilde{\chi}_k^-}^2(s + u_2 - u_7)]
-\frac{1}{su_7}[t_1u_6(s_{5\Delta}
\\ && \hspace{1.5cm}
+ 2m_{\tilde{\chi}_k^-}^2) - u_6^2(u_2 - 2m_{\tilde{\chi}_k^-}^2)
+ t_1u_2m_{\tilde{t}_i}^2 + u_6m_{\tilde{t}_i}^2(2t_2 + u_2)]\}
+\frac{1}{36ss_4}\{\frac{1}{2t'u_7}[2u_6^2(t_2
\\ && \hspace{1.5cm}
+ u_2) - s_{5\Delta}u_6u' + t_2u'm_{\tilde{t}_i}^2]
-\frac{9}{4t_1u'}[(2t_2 + u_2)(u_6(t' -s_{3\Delta} + u_2 + u_7) +
2t'(s_{5\Delta}
\\ && \hspace{1.5cm}
- 2m_{\tilde{t}_i}^2)) -t_2u_7(s_{3\Delta} + 4t_2 + t' + 3u_2 -
u_7)] +\frac{16m_{\tilde{t}_i}^2}{s_4t_1}[u_2(t' + u_6) + t_2(2t'
+ 2u_6
\\ && \hspace{1.5cm}
+ u_7 + u')] -\frac{1}{t_1u_7}[2u_6m_{\tilde{t}_i}^2(2t_2 +u_2) +
t_1(s_{5\Delta}u_6 + u_2m_{\tilde{t}_i}^2)] \}.
\end{eqnarray*}

For the subprocess with two initial-state gluons, we find
\begin{eqnarray*}
\overline{|M^{gg}_{ik}|}^2 = -\frac{3}{8}
\overline{|M^{gb}_{ik}|}^2 (s \leftrightarrow u', \ s_4
\leftrightarrow u_6, \ s_3 \rightarrow s_4 + u_7 +u', \
s_{3\Delta} \rightarrow s_4 + u_7 +u' +\Delta_t).
\end{eqnarray*}

For the subprocess with the initial-state $q\bar{q}$, we find
\begin{eqnarray*}
&&\overline{|M^{q\bar{q}}_{ik}|}^2=\frac{1}{9s^2} X_{ik}
\{\frac{2}{s_5^2}[\Delta_t(t'u_2 + t_2u') + (t'u_7 +
u_6u')(s_{5\Delta} + 2m_{\tilde{\chi}_k^-}^2)]- \frac{1}{s_3^2}[
ss_3(\Delta_s + s_3
\\ && \hspace{1.5cm}
- s_4 - s_{5\Delta}) + s_3(t_2 + t' - u_6)(u_7 -u_2 - u')] -
\frac{1}{s_3s_5}[s_{5\Delta}t'u_2 -2ss_3(\Delta_t + s_{5\Delta})
\\ && \hspace{1.5cm}
- s_4u_2(2t_2 + t' - u_6) +t'u_7(s_3 - s_{5\Delta})- s_3u_7(2u_6
-t_2 ) + s_3u_6(u_2 + u') - s_4t_2(u'
\\ && \hspace{1.5cm}
-u_7) - 2m_{\tilde{\chi}_k^-}^2 ( t'(u_7 -u_2)-2ss_4) + u'(t_2 +
2t' - u_6)(s_{5\Delta} + 2m_{\tilde{\chi}_k^-}^2)] \}.
\end{eqnarray*}
For the subprocess with the initial-state $bq$ or $b\bar{q}$, we
find
\begin{eqnarray*}
&&\overline{|M^{bq(\bar{q})}_{ik}|}^2=
\overline{|M^{q\bar{q}}_{ik}|}^2 (s\leftrightarrow u',\ s_4
\leftrightarrow u_6, \ t_2 \leftrightarrow s_{3\Delta}, \
s_3\rightarrow t_2, \frac{1}{s_3} \rightarrow \frac{1}{t_1} ).
\end{eqnarray*}
For the subprocess with the initial-state $b\bar{b}$, we find
\begin{eqnarray*}
&&\overline{|M^{b\bar{b}}_{ik}|}^2= \frac{1}{27} X_{ik}
\{\frac{3s_{3\Delta}}{s^2s_3^2}[s(s_4 + s_{5\Delta} - \Delta_s -
s_{3\Delta}) + (t_2 + t' - u_6)(u_2 - u_7 +u')]
\\ && \hspace{1.5cm}
+ \frac{2t'}{ss_5^2u'}[s_{5\Delta}u_7 -
u_2m_{\tilde{t}_i}^2 + (u_2 + 2u_7)m_{\tilde{\chi}_k^-}^2] -
\frac{1}{ss_3s_5u'}[\Delta_st'u_2 -ss_{3\Delta}s_{5\Delta}
\\ && \hspace{1.5cm}
+ s_{3\Delta}u_6(u_2 - u_7) + s_4t_2u_7 + (s_{3\Delta} -
s_{5\Delta})t'u_7 + (t_2 + t')(-s_4u_2 + s_{5\Delta}u')
\\ && \hspace{1.5cm}
- \Delta_t(ss_{3\Delta} -
t_2u') + 2(ss_4 + (t' - u_6)u')m_{\tilde{\chi}_k^-}^2]+
\frac{1}{4ss_3t_1u'}[2ss_{3\Delta}(s_{3\Delta} - s_4
\\ && \hspace{1.5cm}
- s_{5\Delta} + t') - t'u_2(2t_2 -2s_4  + t') +
m_{\tilde{\chi}_k^-}^2 (2t_2u' + t'(u_2 - 2u_7 + 4u_p)) + 2(s_4t_2
\\ && \hspace{1.5cm}
-s_{3\Delta}t_2 )(s + u_2 - u_7) + s_{5\Delta}t'(u_7 -2s ) +
m_{\tilde{t}_i}^2 (2t_2u' -t'u_2 )] +
\frac{6}{s^2s_5^2}[\Delta_t(t'u_2 + t_2u')
\\ && \hspace{1.5cm}
+ (t'u_7 + u_6u') (s_{5\Delta} + 2m_{\tilde{\chi}_k^-}^2)] -
\frac{3}{s^2s_3s_5}[s_{5\Delta}t'(u_2 - u_7)
-2ss_{3\Delta}(\Delta_t + s_{5\Delta})
\\ && \hspace{1.5cm}
- s_4u_2(2t_2 + t') + s_{3\Delta}u_7(t' - 2u_6) + (s_{3\Delta} +
s_4)(u_2u_6 + t_2u_7) - s_4t_2u' + s_{5\Delta}u'(t_2
\\ && \hspace{1.5cm}
+ 2t' - u_6) + s_{3\Delta}u_6u' + 2m_{\tilde{\chi}_k^-}^2(2ss_4 +
(t_2 + 2t' - u_6)u' + t'(u_2 - u_7))]\}
\\ && \hspace{1.5cm} + (s \leftrightarrow u', \ s_4 \leftrightarrow u_6,
\ t_2 \leftrightarrow s_{3\Delta}, \ s_3 \leftrightarrow t_1 ).
\end{eqnarray*}
For the subprocess with the initial-state $bb$, we find
\begin{eqnarray*}
&&\overline{|M^{bb}_{ik}|}^2= \overline{|M^{b\bar{b}}_{ik}|}^2
(s\leftrightarrow t',\ s_4 \leftrightarrow u_7, \ u_2
\leftrightarrow s_{3\Delta}, \ s_3\leftrightarrow u_1).
\end{eqnarray*}
%
%
%%%%%%%%%%%%%%%%%%%%% REFERENCES %%%%%%%%%%%%%%%%%%%%%%%%%%%%%%%%%%%%%%%%%%%%%%

%%%%%%%%%%%%%%%%%%%%%% tree-level diagrams %%%%%%%%%%%%%%%%%%%%%%%%
\newpage
\vspace{3.0cm}
%%%%%%%%%%%%%%%%%% one-loop correction diagrams %%%%%%%%%%%%%%%%
\newpage
%%%%%%%%%%%%%%%%%%%%% box diagrams %%%%%%%%%%%%%%%%%%%%%%%%%%%%
%%%%%%%%%%%%%%%%% real correction diagrams %%%%%%%%%%%%%%%%%%%%%%%
\begin{figure}
\centerline{\epsfig{file=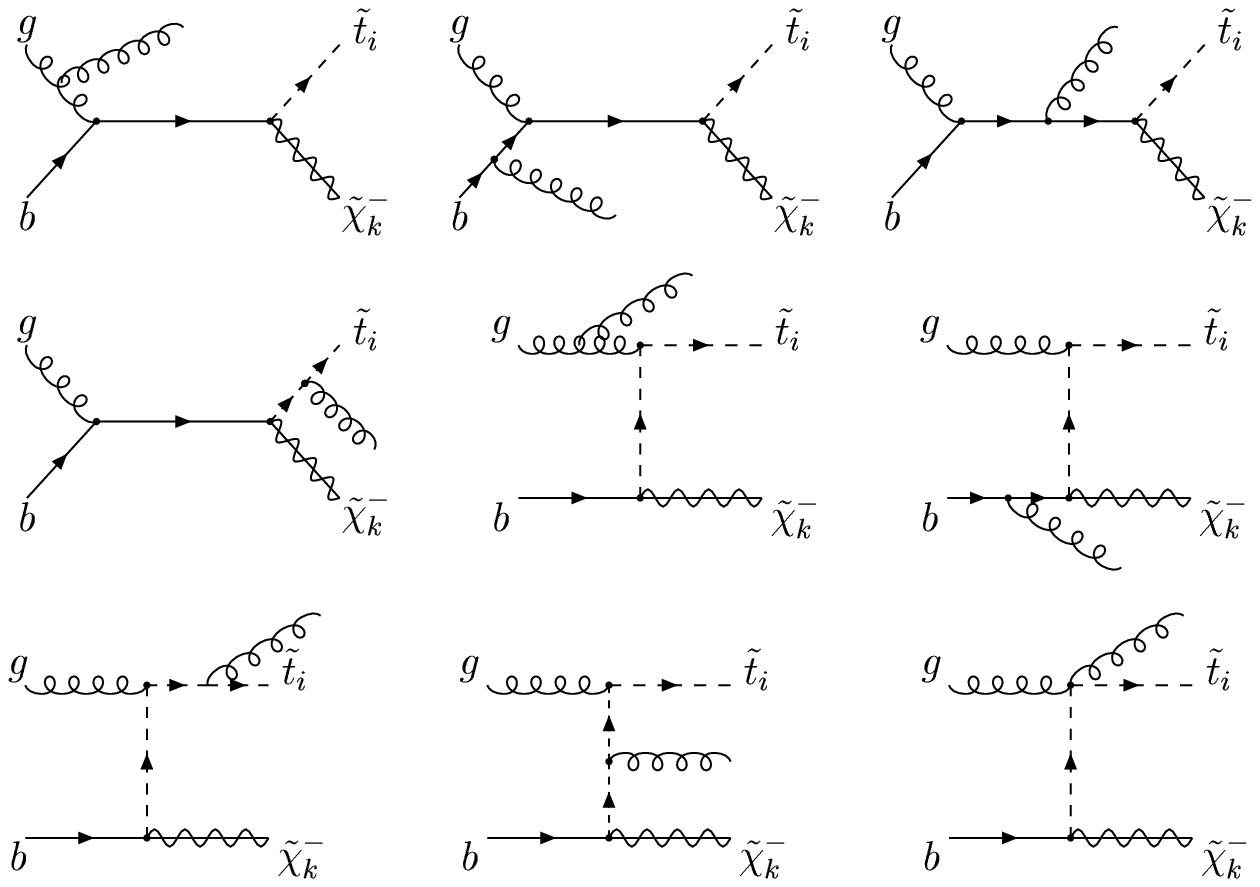, width=350pt}}
\caption[]{Feynman diagrams for the real gluon emission
contributions. \label{fynreal}}
\end{figure}
%%%%%%%%%%%%%%%%%%%%%%% splitting diagrams %%%%%%%%%%%%%%%%%%%%%%%
\begin{figure}
\centerline{\epsfig{file=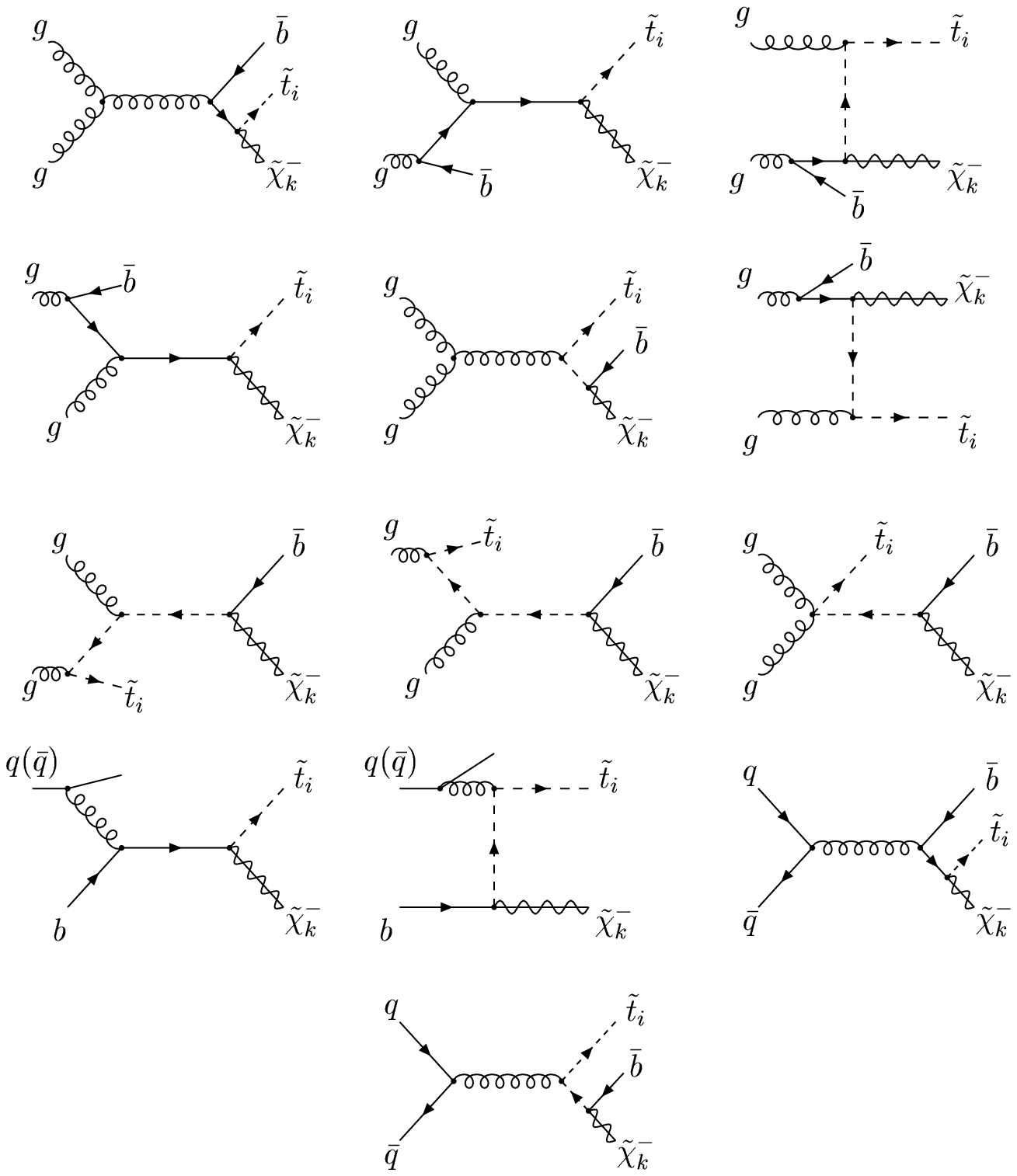, width=350pt}}
\caption[]{Feynman diagrams for the emission of a massless
(anti)quark. Here $q=u,d,c,s,b$. \label{splitting}}
\end{figure}

\begin{figure}
\centerline{\epsfig{file=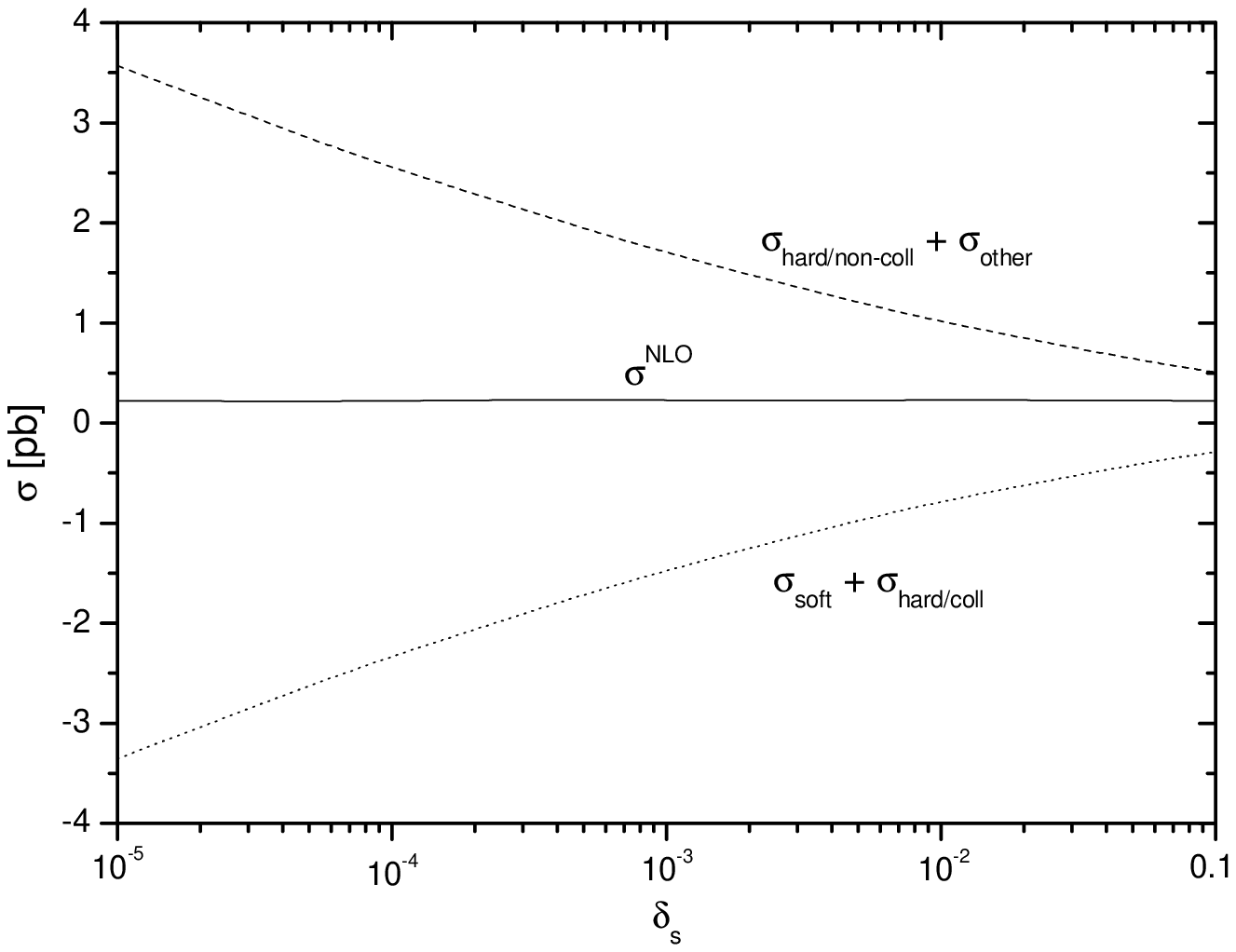,width=400pt}}
\caption[]{Dependence of the total cross sections for the
$\tilde{t}_1\tilde{\chi}_1^-$ production at the LHC on the cutoff
$\delta_s$, assuming $\mu=-200$ GeV, $M_2=300$ GeV,
$m_{\tilde{t}_1}=250$ GeV, $\tan\beta=30$ and
$\delta_c=\delta_s/100$. \label{deltaseps}}
\end{figure}

\begin{figure}
\centerline{\epsfig{file=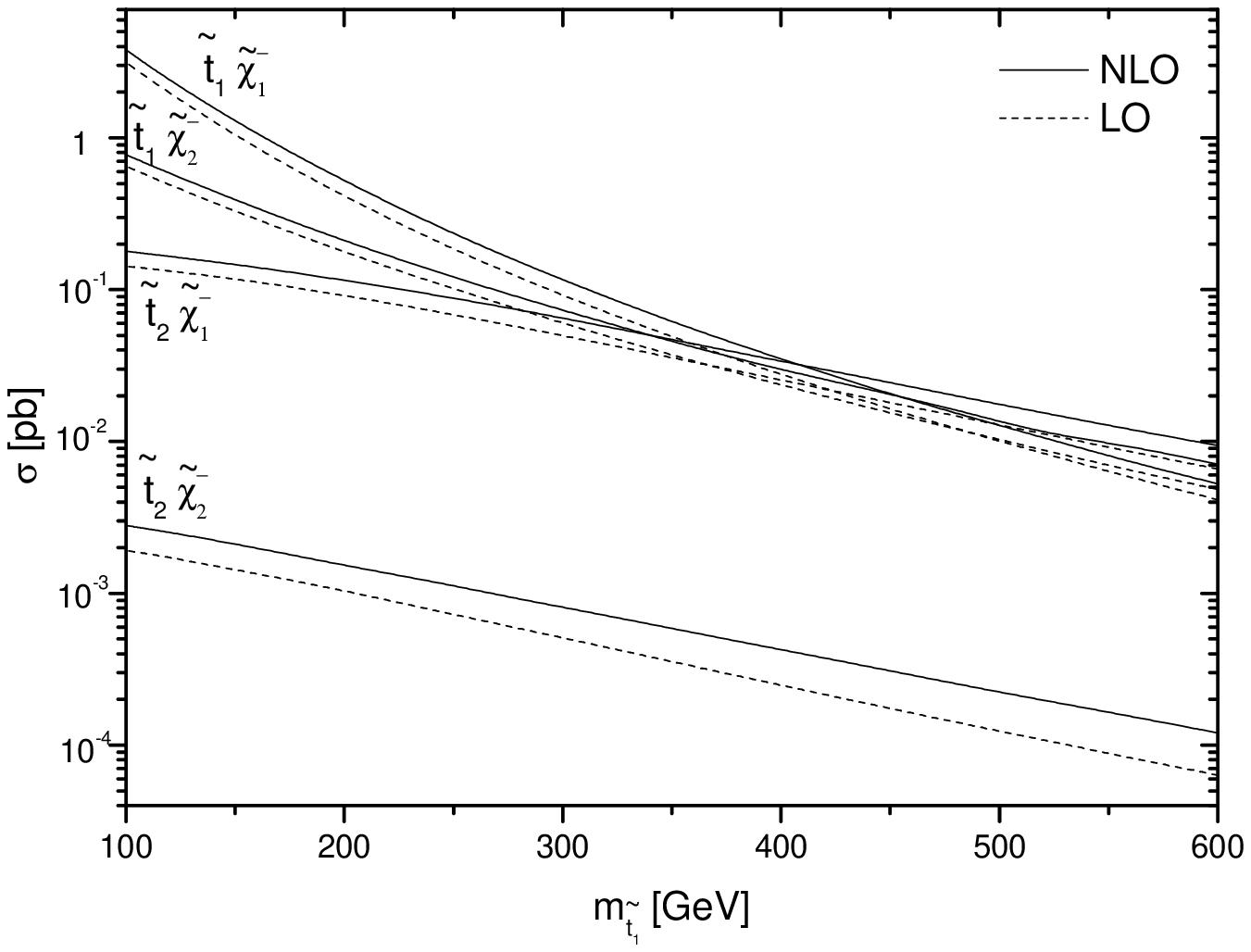, width=400pt}}
\caption[]{Dependence of the total cross sections on
$m_{\tilde{t}_1}$ for the $\tilde{t}_i\tilde{\chi}_k^-$
productions at the LHC, assuming $\mu=-200$ GeV, $M_2=300$ GeV and
$\tan\beta=30$. \label{msteps}}
\end{figure}

\begin{figure}
\centerline{\epsfig{file=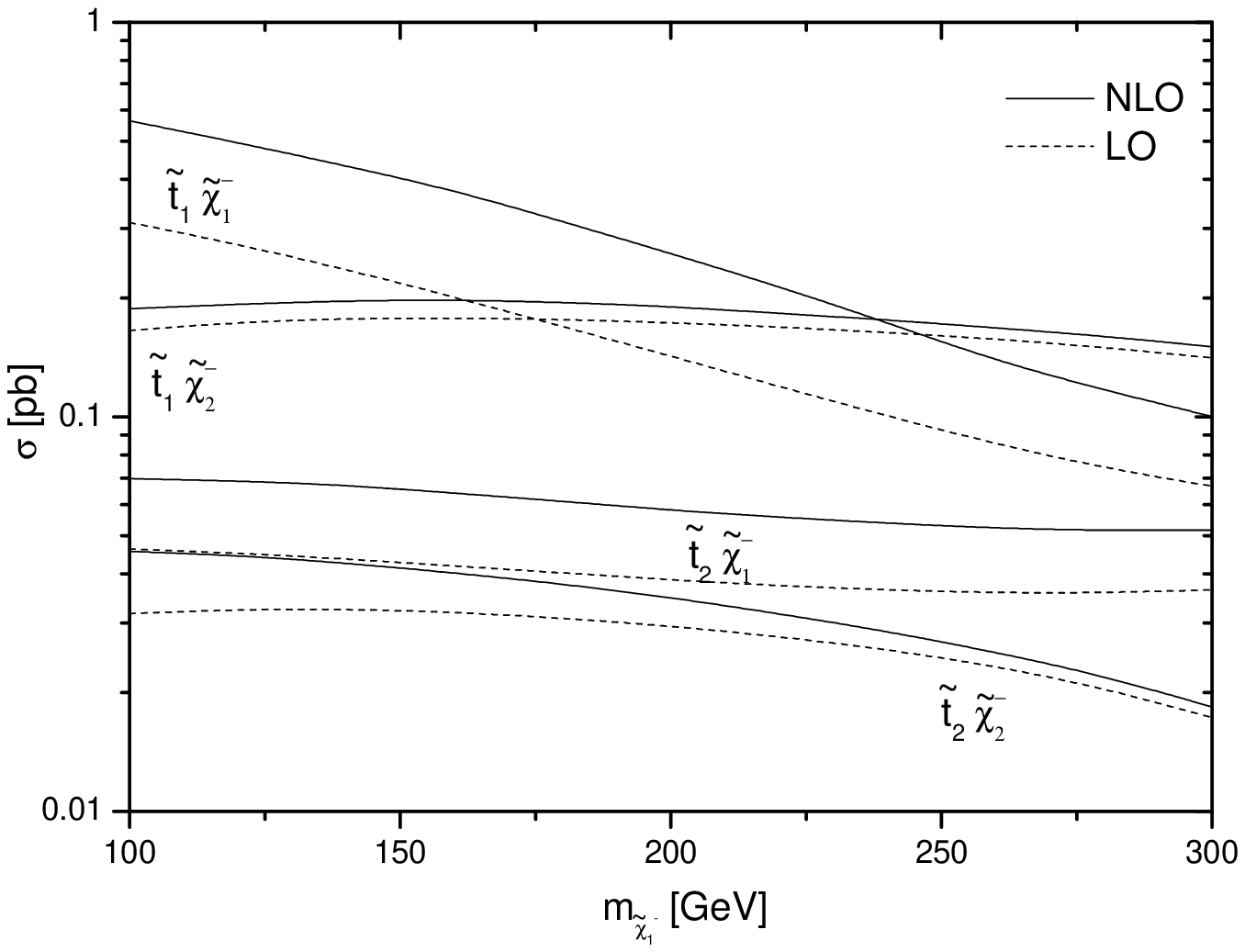, width=400pt}}
\caption[]{Dependence of the total cross sections on
$m_{\tilde{\chi}_1^-}$ for the $\tilde{t}_i\tilde{\chi}_k^-$
productions at the LHC, assuming $\mu=-400$ GeV,
$m_{\tilde{t}_1}=250$ GeV and $\tan\beta=30$. \label{mxpeps}}
\end{figure}

\begin{figure}
\centerline{\epsfig{file=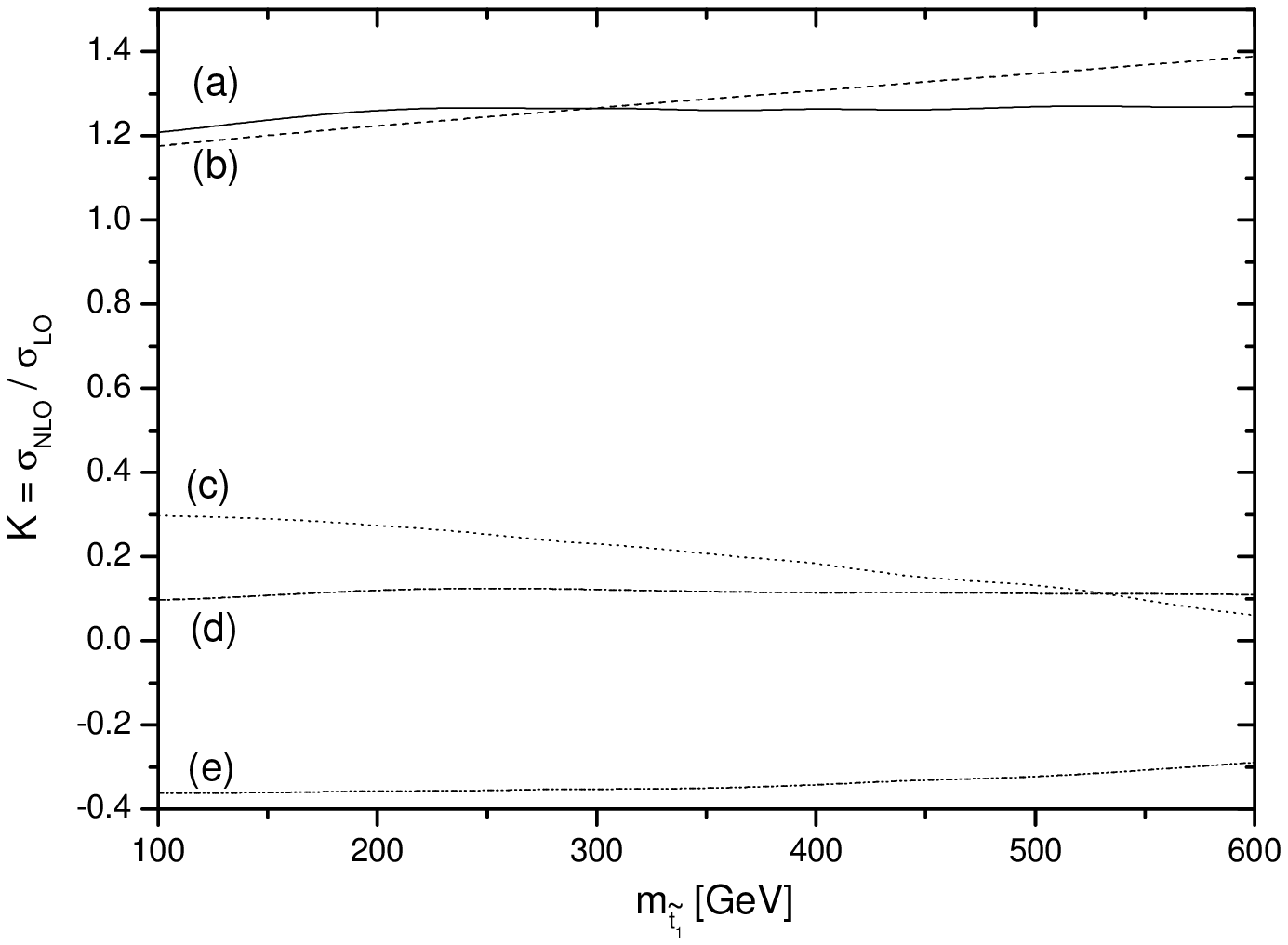, width=400pt}} \caption[]{$K=
\sigma_{NLO}/\sigma_{LO}$ factor for the
$\tilde{t}_1\tilde{\chi}_1^-$ production at the LHC as a function
of $m_{\tilde{t}_1}$, assuming $\mu=-200$ GeV, $M_2=300$ GeV and
$\tan\beta=30$. $\sigma_{NLO}$ is the NLO total cross section for
$(a)$, the improved Born cross section for $(b)$, the corrections
of the subprocess with two initial-state gluons for $(c)$, the
corrections of the massless (anti)quark emission subprocess
Eq.(\ref{qb})-(\ref{qbq}) for $(d)$, and the virtual and real
gluon emission corrections for $(e)$. \label{ratioeps}}
\end{figure}

\begin{figure}
\centerline{\epsfig{file=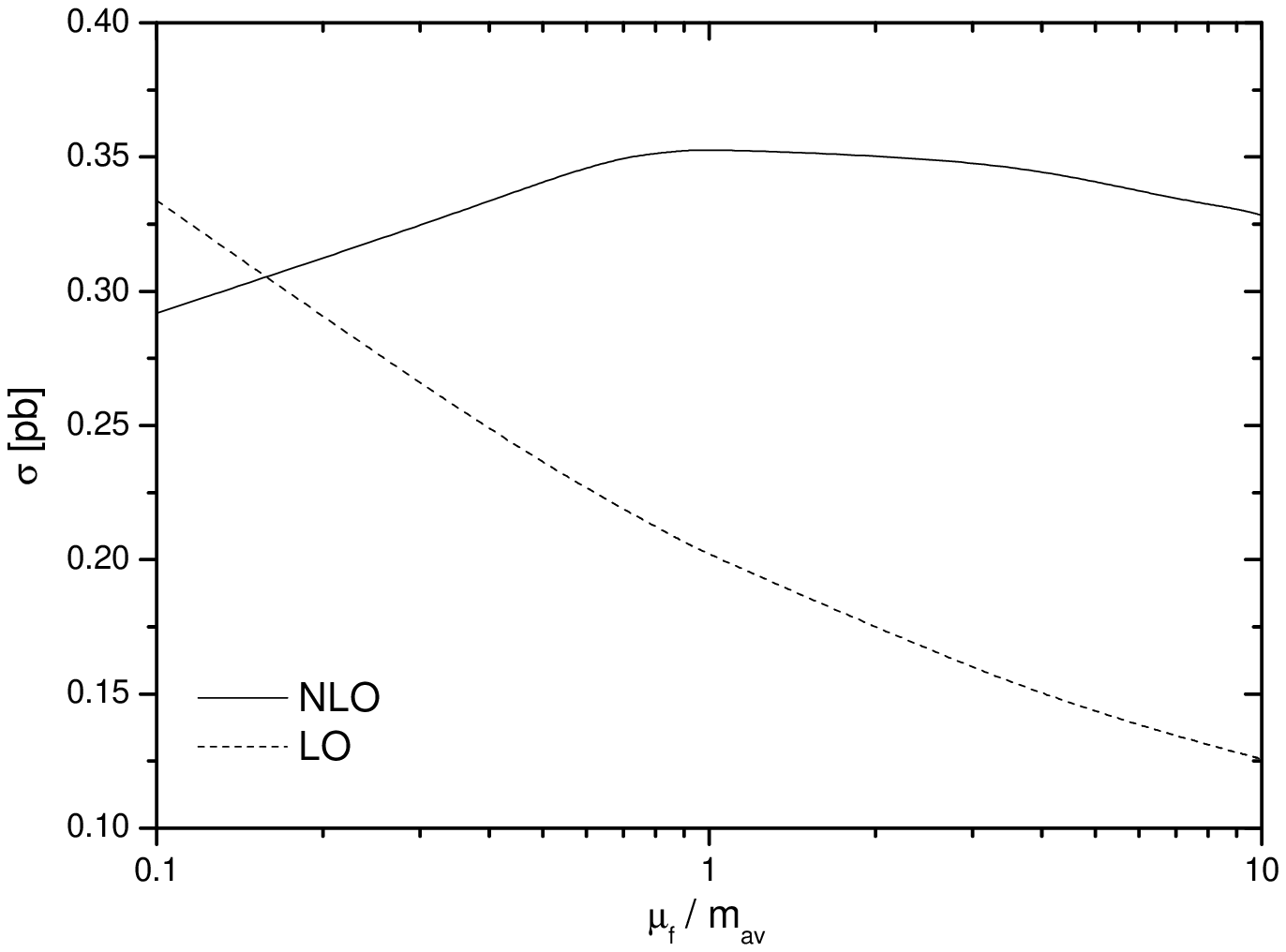, width=400pt}}
\caption[]{Dependence of the total cross sections for the
$\tilde{t}_1\tilde{\chi}_1^-$ production at the LHC on the
renormalization/factorization scale, assuming $\mu=-200$ GeV,
$M_2=300$ GeV, $\tan\beta=30$, $m_{\tilde{t}_1}=250$ GeV,
$\mu_r=\mu_f$ and $m_{\rm av}=(m_{\tilde{t}_1}
+m_{\tilde{\chi}_1^-})/2$. \label{mumu0eps}}
\end{figure}

\begin{figure}
\centerline{\epsfig{file=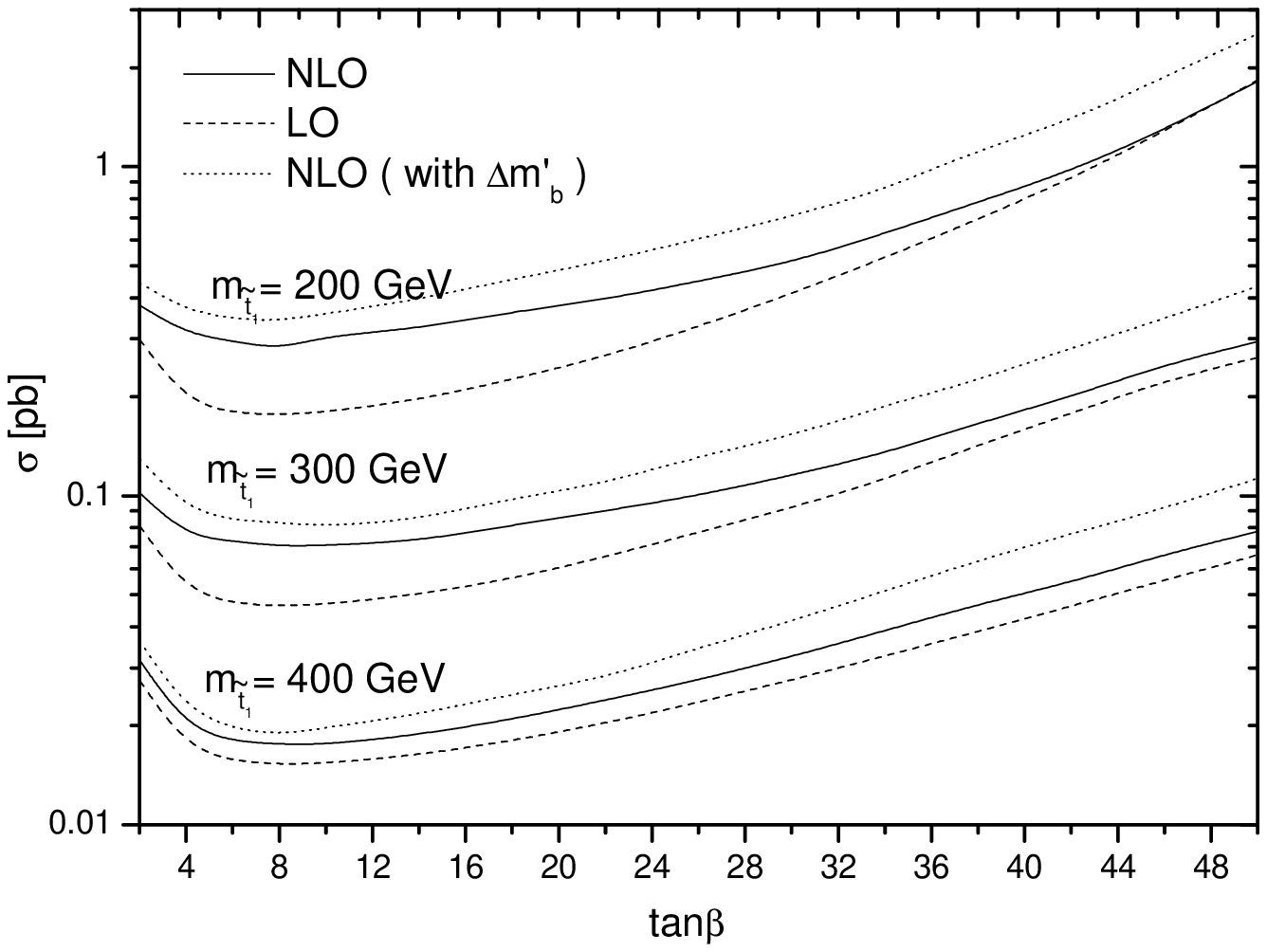, width=400pt}}
\caption[]{Dependence of the total cross sections for the
$\tilde{t}_1\tilde{\chi}_1^-$ production at the LHC on the
parameter $\tan\beta$, assuming $\mu=-200$ GeV and $M_2=300$ GeV.
\label{taneps}}
\end{figure}

\begin{figure}
\centerline{\epsfig{file=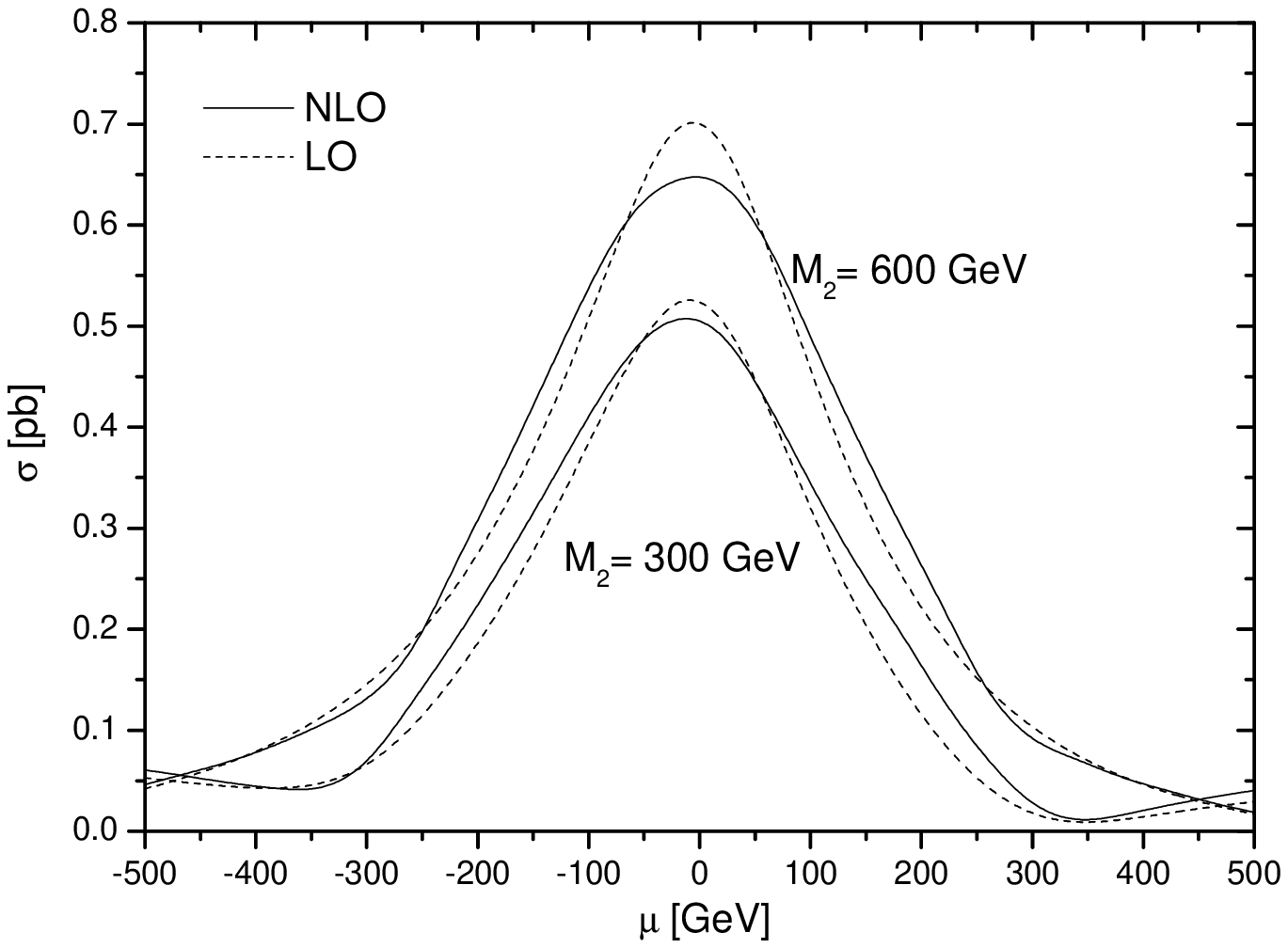, width=400pt}}
\caption[]{Dependence of the total cross sections for the
$\tilde{t}_1\tilde{\chi}_1^-$ production at the LHC on the
parameter $\mu$, assuming $m_{\tilde{t}_1}=250$ GeV and
$\tan\beta=30$. \label{mueps}}
\end{figure}

\begin{figure}
\centerline{\epsfig{file=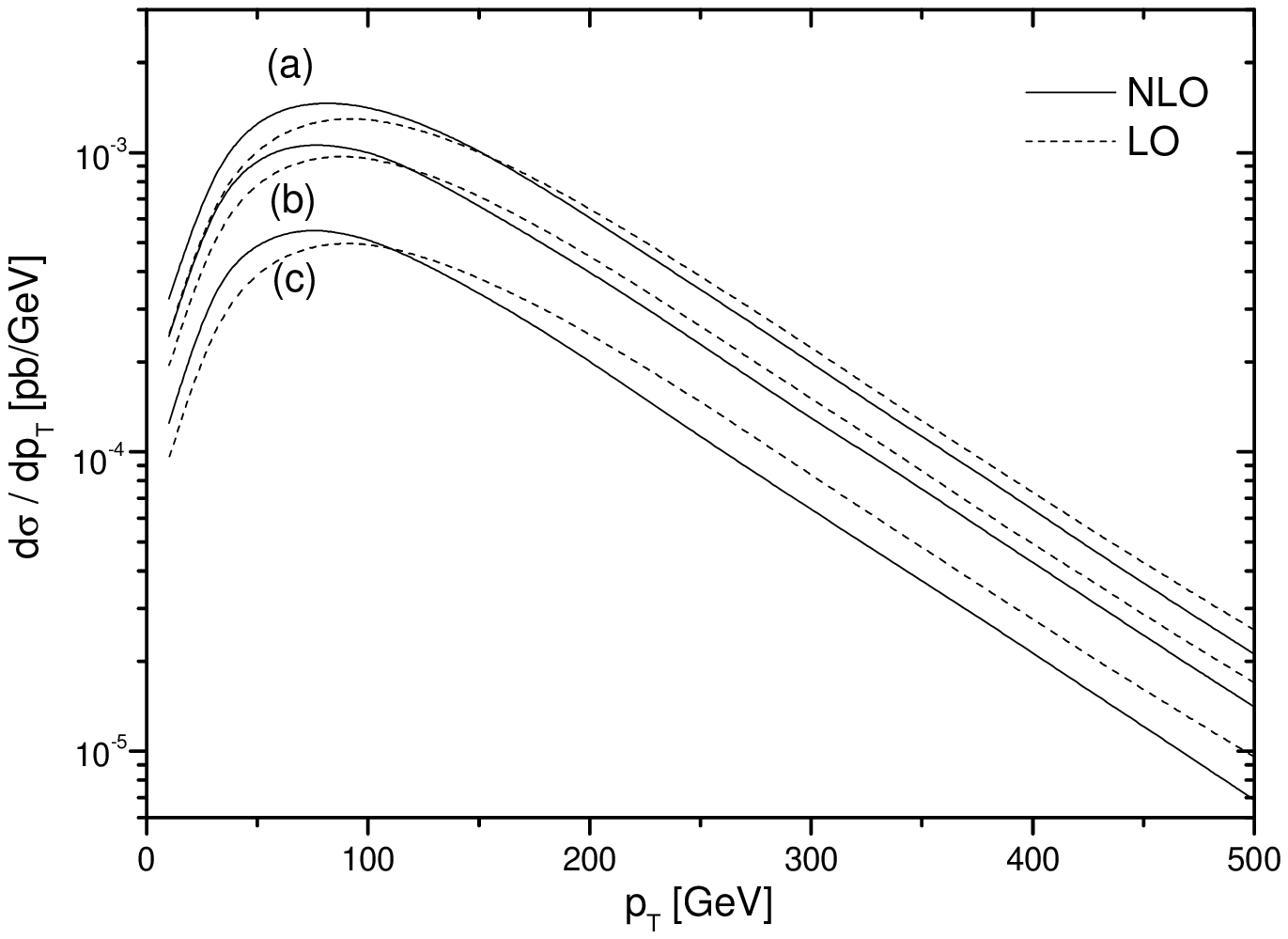, width=400pt}}
\caption[]{Differential cross sections in the transverse momentum
$p_T$ of $\tilde{\chi}_1^-$ for the $\tilde{t}_1\tilde{\chi}_1^-$
production at the LHC, assuming $\mu=-200$ GeV and
$m_{\tilde{t}_1}=250$ GeV. For $(a)$, $\tan\beta=30$ and $M_2=600$
GeV; for $(b)$, $\tan\beta=30$ and $M_2=300$ GeV; for $(c)$,
$\tan\beta=4$ and $M_2=300$ GeV. \label{pteps}}
\end{figure}

\end{document}